\documentclass{article}
\usepackage{arxiv}

\pdfoutput=1

\usepackage{braket}
\usepackage{amssymb}
\usepackage{mathtools}
\usepackage{amsmath}
\usepackage{mathdots}

\usepackage[title]{appendix}
\usepackage{ytableau}
\usepackage{tikzit}
\tikzstyle{red dot}=[fill=red, draw=black, shape=circle]
\tikzstyle{green dot}=[fill=green, draw=black, shape=circle]
\tikzstyle{control dot}=[fill=black, draw=black, shape=circle]
\tikzstyle{medium box}=[fill=white, draw=black, shape=rectangle, minimum width=1.5cm, minimum height=1cm]
\tikzstyle{small box}=[fill=white, draw=black, shape=rectangle, minimum height=0.5cm, minimum width=0.5cm]
\tikzstyle{Hadamard}=[fill=yellow, draw=black, shape=rectangle]
\tikzstyle{weyl}=[fill=white, draw={rgb,255: red,0; green,0; blue,109}, shape=rectangle, minimum height=0.5cm, minimum width=0.5cm]

\tikzstyle{dashes}=[-, dashed, draw={rgb,255: red,191; green,191; blue,191}, dash pattern=on 2mm off 1mm]

\begin{document}

\title{
	%Title TBC
	%Exploring an application of the Schur--Weyl duality using Okounkov--Vershik theory
	%A multigraph approach for performing the Quantum Schur Transform inspired by Okounkov--Vershik theory
	A Multigraph Approach for Performing\\ 
	the Quantum Schur Transform 
	%An adaptation of Okounkov--Vershik's representation theory of the symmetric groups to the Schur--Weyl duality
}

\author{Edward Pearce--Crump\\
Imperial College London\\
\texttt{ep1011@ic.ac.uk} \\
}

\date{\today}
\maketitle

\begin{abstract}
	%We present an implementation of the Quantum Schur Transform that is founded upon an adaptation of the Okounkov--Vershik formulation of the representation theory of the symmetric groups to the Schur--Weyl duality.
	%We present a different theoretical approach to the one that results in the well-known implementation of the Quantum Schur Transform \textbf{CITE} which is founded upon an adaptation of the Okounkov--Vershik formulation of the representation theory of the symmetric groups to the Schur--Weyl duality.
	We take inspiration from the Okounkov--Vershik approach to the representation theory of the symmetric groups to develop a new way of understanding how the Schur--Weyl duality can be used to perform the Quantum Schur Transform.
	The Quantum Schur Transform is a unitary change of basis transformation between the computational basis of $(\mathbb{C}^d)^{\otimes n}$ and the Schur--Weyl basis of $(\mathbb{C}^d)^{\otimes n}$. 
	We describe a new multigraph, which we call the Schur--Weyl--Young graph, that 
	%is a modified version of the Young graph for the symmetric group $S_n$, 
	represents both standard Weyl tableaux and standard Young tableaux in the same diagram.
	We suggest a major improvement on Louck's formula for calculating the transition amplitudes between two standard Weyl tableaux appearing in adjacent levels of the Schur--Weyl--Young graph for the case $d=2$, merely by looking at the entries in the two tableaux.
	The key theoretical component that underpins our results is the discovery of a branching rule for the Schur--Weyl states, which we call the Schur--Weyl branching rule.
	This branching rule allows us to perform the change of basis transformation described above in a straightforward
	manner for any $n$ and $d$.
	%This paper contains four new contributions. 
	%Firstly, a modified version of the Young diagram for the symmetric group $S_n$, which we call the Schur--Weyl--Young diagram, representing both standard Young tableaux and standard Weyl tableaux in the same diagram. 
	%Secondly, we suggest a major improvement on Louck's formula for calculating the transition amplitudes between two Weyl tableaux of sizes $n$ and $n+1$ for $d=2$, merely by looking at the entries of the two tableaux.
	%Thirdly, we describe a new branching rule, which we call the Schur--Weyl Branching Rule, that allows us to easily change basis between the computational basis and the Schur--Weyl basis of $(\mathbb{C}^d)^{\otimes n}$.
	%Fourthly, we use this branching rule to provide a linear time implementation of the Quantum Schur Transform for any $n, d$, giving a quantum circuit for the case when $d=2$.
\end{abstract}

\section{Introduction}
%The Schur--Weyl duality is a famous result in representation theory relating, for any positive integers $n, d$, irreducible representations of the symmetric group $S_n$ with irreducible representations of the special unitary group $SU(d)$.
The Schur--Weyl duality is a famous result in representation theory relating, for any positive integer $d$ and any non-negative integer $n$, the irreducible representations appearing in the decomposition of the $n$-fold tensor product space $(\mathbb{C}^d)^{\otimes n}$ considered both as a representation of
the symmetric group $S_n$, the group of permutations of $n$ objects, and as a representation of the special unitary group $SU(d)$, the Lie group of $d \times d$ unitary matrices with determinant one.

This result 
%underpins the existence of
provides the key theoretical foundation for the existence of 
%The Schur--Weyl duality is the theoretical result which underpins the existence of 
a unitary operator in quantum computing known as the Quantum Schur Transform.
This operator is defined to be the unitary change of basis transformation which maps a quantum state expressed in the computational basis of $(\mathbb{C}^d)^{\otimes n}$ to the same quantum state expressed in the so--called Schur--Weyl basis of $(\mathbb{C}^d)^{\otimes n}$.
Considerable effort has been devoted to constructing
%Much effort has recently gone into constructing 
an efficient implementation of the Quantum Schur Transform.
We review the history behind these efforts in Section \ref{relatedwork}.
Notably, all of the implementations that have been discovered so far depend on a choice of either employing the representation theory of the symmetric groups or the representation theory of the unitary groups.
In our approach, we make heavy use of the representation theory of the symmetric groups.

The representation theory of the symmetric groups has a long and rich history.
%The approach that is dominant in the literature
The original version of the theory
%was formulated by 
is credited to Young \cite{young}, who made his contributions in the early 1900s, although further, significant contributions were later made by luminaries such as von Neumann \cite{vonNeumann}, Weyl \cite{weyl} and James \cite{james}, amongst others. 
Young constructed the irreducible representations of the symmetric groups by introducing a class of diagrams originally called tableaux, but which are now commonly referred to as Young tableaux.
%which consist of rows of boxes corresponding to a partition of $n$ that is filled with entries from $\{1, \dots, n\}$ such that each integer appears once and only once in the diagram.
In taking a combinatorial approach to these tableaux, 
%Young proved that there exists a unique irreducible representation (up to equivalence) of the symmetric group $S_n$ associated to each partition of $n$. 
Young proved that the only irreducible representations of the symmetric group $S_n$ (up to equivalence) are in bijective correspondence with the partitions of $n$.
%He achieved this by forming a subgroup associated to each partition called the Young subgroup, and by considering the trivial and sign representations of this subgroup and inducing them up to become representations of the symmetric group $S_n$. 
%The desired irreducible representation arose as the unique common component of the decomposition into irreducibles of the two induced representations described.
It was shown further that a branching rule exists for the irreducible representations of the symmetric groups.
%Namely, this rule states that 
%an irreducible of $S_n$, when restricted to be a representation of $S_{n-1}$, could be expressed as a direct sum of irreducibles $S_{n-1}$ with multiplicities equal to either zero or one.
%Furthermore, the irreducibles appearing with multiplicity one originated from the partition (associated to the irreducible of $S_n$) with a box removed from the partition's Young frame such that the new Young frame corresponded to a valid partition of $n-1$. 
These irreducible representations can be represented in a graphical structure commonly known as the Young graph, consisting of levels indexed by the non-negative integers.
%The vertices appearing in each level $n$ are the partitions of $n$, representing the equivalence classes of irreducibles of $S_n$, with edges appearing between partitions in adjacent levels $n-1$ and $n$ if the irreducible in level $n-1$ appears with multiplicity one in the decomposition of the irreducible in level $n$ when restricted to $S_{n-1}$.
The branching rule and the Young graph are major components of the theory; yet they only appear as a corollary at the end of Young's formulation of the theory. 

%the irreducibles found being the partitions edges appearing 
\newpage

In 1996, a groundbreaking paper titled 
%``
\textit{A New Approach to the Representation Theory of the Symmetric Groups}
%" 
was published in Russian by Okounkov and Vershik \cite{okounkov-russian}.
Its translation into English appeared in 2005 in the Journal of Mathematical Sciences \cite{okounkov-english}.
Okounkov and Vershik took a rather different approach to Young.
They saw that the symmetric groups naturally form a chain of groups, namely that $S_{n-1}$ is a subgroup of $S_n$, and they recognised that the branching rule and the graphical approach to representing the irreducible representations of this chain of groups was central to the theory.
Hence they made this their starting point, and consequently only derived, as an ``auxiliary element of the construction" \cite{okounkov-english}, the Young tableaux.
The authors suggest that their approach is the more natural one, given that the important combinatorial objects of the theory arise in a natural way, unlike in Young's formulation, where they are introduced at the beginning without any motivation for their introduction.

In this paper, we take inspiration from the Okounkov--Vershik approach to the representation theory of the symmetric groups to develop a new way of understanding 
how the Schur--Weyl duality can be used to implement the Quantum Schur Transform.
%the Schur--Weyl duality, and in particular, the Quantum Schur Transform.
We define a new multigraph, which we call the Schur--Weyl--Young graph, to give a graphical depiction of the Schur--Weyl states. 
We also derive a branching rule for the Schur--Weyl states of $(\mathbb{C}^d)^{\otimes n}$ using this graph, which we call the Schur--Weyl branching rule, and 
%consequently show how the Quantum Schur Transform can be implemented on a quantum computer consisting of $n$ qudits in \textbf{WHAT TIME?} using this branching rule.
consequently describe a procedure for performing the Quantum Schur Transform 
for any $n$ qudits 
in time polynomial in $n$ and $d$ using this branching rule.
In addition, we show how we can dramatically speed up the calculation of the amplitudes involved in the Schur--Weyl branching rule for the case $d=2$ merely by counting certain entries in the pairs of the Weyl tableaux that are involved.

This paper is organised as follows. 
%In Section \ref{relatedwork}, we review the work existing in the literature that is related to ours. 
In Section \ref{background}, we introduce the required background material from representation theory, 
namely the two major approaches to the representation theory of the symmetric group -- the first formulated by Young, the second by Okounkov and Vershik -- together with particular irreducible representations of $SU(d)$ that are of interest to us, before finishing with a description of the Schur--Weyl duality and the Schur--Weyl basis.
In Section \ref{mainresults}, we present our main theoretical results; in particular, we introduce the Schur--Weyl--Young graph and show how it leads to a branching rule for Schur--Weyl states.
In Section \ref{patternrulessection}, we present a simple set of rules for calculating the transition amplitudes between two standard Weyl tableaux appearing in adjacent levels of the Schur--Weyl--Young graph for the case $d=2$, with a proof given in Appendix \ref{patternproof}.
We use these rules to give examples of the Schur--Weyl branching rule in Section \ref{examplesswbranchingrule}.
We present our procedure for performing the Quantum Schur Transform in Section \ref{schurtransform}.
%that uses the ideas presented in the previous sections.
In Section \ref{relatedwork}, we review the work existing in the literature that is related to ours, before concluding in Section \ref{conclusion}.

\section{Preliminaries} \label{background}

\subsection{Remark on Terminology Used}
In the literature, there are different vocabularies for describing the same ideas depending on whether one is a mathematician or a physicist.
We aim, where possible, to use the mathematician's vocabulary (tableaux, contents etc.), 
noting that it would not take much of a leap to convert these ideas into the physicist's vocabulary (spin labels, spin components etc.).
Furthermore, all representations in the following are considered to be over the field of complex numbers.

\subsection{An Introduction to the Representation Theory of the Symmetric Groups}

In this section, we introduce the key terminology and the major results from each of the formulations of the representation theory of the symmetric groups that were discussed in the Introduction.
For more details on the Young formulation, see \cite{sagan}, and for more details on the Okounkov--Vershik formulation, see \cite{tolli}, \cite{okounkov-english}.

\subsubsection{The Young Formulation}

The Young formulation uses a combinatorial approach to develop the theory, beginning with partitions of a non-negative integer $n$ and associating to each one a so-called Young frame.

A partition $\lambda$ of $n$, denoted by $\lambda \vdash n$, is defined to be a tuple $(\lambda_1, \lambda_2, \dots, \lambda_k)$ such that $\lambda_1 \geq \lambda_2 \geq \ldots \geq \lambda_k \geq 0$, $\sum_{i = 1}^{k} \lambda_i = n$ and $\lambda_i \in \mathbb{Z}$ for all $i$.
We say that such a partition consists of $k$ parts, and we define its length to be $k$.
Note that in this definition, we allow partitions with zero parts, that is, some of the $\lambda_i$ can be $0$.

\begin{figure}[ht]
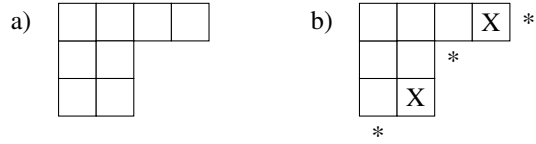

  \ctikzfig{youngframe}
	\caption{a) The Young frame for the partition $(4, 2, 2, 0) \vdash 8$.
	b) X denotes the removable boxes, whereas * denotes the addable boxes.}
  \label{youngframe}
\end{figure}

To every partition $\lambda \vdash n$ we associate a Young frame of shape $\lambda$, that is, a diagram consisting of $k$ rows, where row $i$ consists of $\lambda_i$ boxes (by convention, the row number increases downwards).
We will denote a Young frame of shape $\lambda$ also by $\lambda$, with the context making clear what is being referred to.
Figure \ref{youngframe}a) shows the Young frame for the partition $(4, 2, 2, 0) \vdash 8$.

For each partition $\lambda \vdash n$ we can assign to its corresponding Young frame of shape $\lambda$ a set of coordinates, with the rows being numbered from top to bottom (with starting index $1$) and the columns from left to right (also with starting index $1$).
We say that the box at coordinates $(i, j)$ is \textit{removable} if there is no box at the locations $(i + 1, j)$ and $(i, j+1)$.
We say that the box at coordinates $(i, j)$ is \textit{addable} if, when $i = 1$, $j = \lambda_1 + 1$, else, when $i > 1$, $\lambda_i = j - 1 < \lambda_{i-1}$.
Figure \ref{youngframe}b) shows the removable (X) and addable (*) boxes for the Young frame of shape $(4, 2, 2, 0) \vdash 8$.

\begin{figure}[ht]
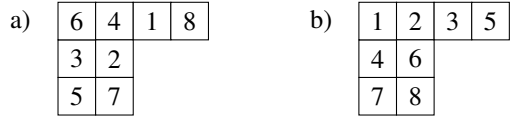

  \ctikzfig{youngtableaux}
  \caption{a) A Young tableau for the partition $(4, 2, 2, 0) \vdash 8$. It is not standard.
	   b) A standard Young tableau for the partition $(4, 2, 2, 0) \vdash 8$.}
  \label{youngtableaux}
\end{figure}

For each partition $\lambda \vdash n$, a Young tableau of shape $\lambda$ is a bijection between the set $\{1, 2, \ldots, n\}$ and the boxes of the Young frame of shape $\lambda$.
We represent such a bijection diagrammatically by filling in the boxes of the Young frame with the numbers $1, 2, \ldots, n$ such that each number appears in exactly one box. 
Figure \ref{youngtableaux}a) gives an example of a Young tableau for the partition $(4, 2, 2, 0) \vdash 8$.
A Young tableau of shape $\lambda$ is said to be standard if the numbers are increasing in each row (from left to right) and in each column (from top to bottom).
Figure \ref{youngtableaux}b) gives an example of a standard Young tableau for the partition $(4, 2, 2, 0) \vdash 8$.
Note that the Young tableau given in Figure \ref{youngtableaux}a) is not standard.

Now for each partition $\lambda = (\lambda_1, \lambda_2, \dots, \lambda_k)\vdash n$, we obtain the following subgroup of $S_n$, namely 
$$
S_{\lambda_1} \times S_{\lambda_2} \times \dots \times S_{\lambda_k}
$$
This is called the Young subgroup for $\lambda$, and we denote it by $S_\lambda$.
By taking the trivial representation of $S_\lambda$ and inducing it up to $S_n$, we obtain a representation of $S_n$ which we denote by $M^{\lambda}$. Notice that the representation is labelled by the partition.
It is called the permutation module for $\lambda$, and it has a basis consisting of the cosets of $S_{\lambda}$ in $S_n$ (there is an alternative formulation in terms of so-called $\lambda$-tabloids~-- see \cite{sagan} for more details.)

$M^\lambda$ itself may or may not be irreducible, but it can be shown that, as part of its decomposition into irreducible representations of $S_n$, it has one copy of an irreducible representation which is denoted by $S^\lambda$.
$S^\lambda$ is called a Specht module, and it can be shown that the set
$$
\{S^\lambda \mid \lambda \vdash n\}
$$
is a complete set of irreducible representations for $S_n$ (up to equivalence). 
That is, the irreducible representations of $S_n$ (up to equivalence) are in bijective correspondence with the set of partitions of $n$.
It can further be shown that each $S^\lambda$ has a basis which is in bijective correspondence with the set of standard Young tableaux of shape $\lambda$.

It is well known that there is a branching rule for irreducible representations of $S_n$.
%that is intimately connected to the Young diagram.
It exists in two versions, depending on whether we are restricting or inducing representations of $S_n$.
The restriction version states that for any partition $\lambda \vdash n$, its accompanying irreducible representation of $S_n$, $S^\lambda$, can be expressed in the following form when restricted to be a representation of $S_{n-1}$:
\begin{equation} \label{youngrestriction}
	S^\lambda \downarrow_{S_{n-1}}^{S_n} = \bigoplus_{\mu \vdash n-1: \mu \leftarrow \lambda} S^\mu	
\end{equation}
where the direct sum is taken over all partitions $\mu \vdash n-1$ such that $\mu$ can be obtained by removing a box from the Young frame of $\lambda$ (which is denoted by $\mu \leftarrow \lambda$).
%there is an arrow connecting $\lambda$ and $\mu$ in the Young diagram.

The induction version states that for any partition $\lambda \vdash n$, its accompanying irreducible representation of $S_n$, $S^\lambda$, can be expressed in the following form when induced to be a representation of $S_{n+1}$:
\begin{equation}
	S^\lambda \uparrow_{S_n}^{S_{n+1}} = \bigoplus_{\mu \vdash n+1: \lambda \leftarrow \mu} S^\mu	
\end{equation}
where the direct sum is taken over all partitions $\mu \vdash n+1$ such that $\mu$ can be obtained by adding a box to the Young frame of $\lambda$ (which is denoted by $\lambda \leftarrow \mu$).
These versions are, in fact, equivalent under the Frobenius Reciprocity Theorem. 

The branching rule for irreducible representations of $S_n$ leads to a way of representing these irreducible representations in terms of a graph, which is known as the Young graph.
Let $P(n)$ be the set of Young frames associated with the partitions of $n$, that is, $P(n) = \{\lambda \mid \lambda \vdash n\}$.

The Young graph has a vertex set which is the disjoint union
$$
\bigcup_{n \geq 0} P(n)
$$
where the set $P(n)$ is called the $n^{\text{th}}$ level of the graph. 
We denote by $\varnothing$ the unique element of $P(0)$.

\begin{figure}[ht]
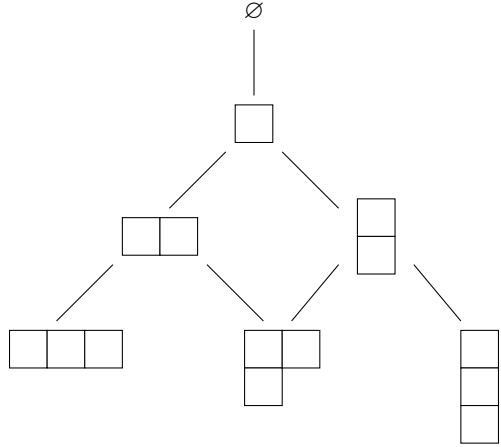

  \ctikzfig{youngdiagram}
  \caption{The Young graph for the symmetric group $S_n$ up to and including $n = 3$.}
  \label{youngdiagram}
\end{figure}

We say that two vertices $\mu \in P(n-1)$ and $\lambda \in P(n)$ are joined by a single edge if and only if
$$
\text{dim Hom}_{S_{n-1}}(S^\mu, S^\lambda \downarrow_{S_{n-1}}^{S_{n}})
= 1
$$
that is, if and only if the multiplicity of $S^\mu$ in the restriction of $S^\lambda$ from $S_n$ to $S_{n-1}$ is $1$.
This is a direct consequence of the branching rule given in (\ref{youngrestriction}).
Figure \ref{youngdiagram} shows the Young graph for the symmetric group $S_n$ up to and including $n = 3$.

% - idea up to here: build it up to the Young graph: partitions, young diagrams, permutation modules, specht modules, branching rule, Young graph$

\subsubsection{The Okounkov--Vershik Formulation}

We noted earlier that the Okounkov--Vershik formulation 
%of the Representation Theory of the Symmetric Groups 
takes a rather different approach to the Young formulation.

Okounkov and Vershik start by looking at the inductive chain of finite groups
\begin{equation} \label{chaingroups}
%\{1\} = S_0 \subset S_1 \subset S_2 \subset \dots \subset S_n \subset \dots
\{1\} = G(0) \subset G(1) \subset G(2) \subset \dots \subset G(n) \subset \dots
\end{equation}
and consider, for each $n$, the set of equivalence classes of irreducible representations of $G(n)$, which is denoted by $G(n)^{\wedge}$. 

From these sets, they form a multigraph called the Bratelli diagram.
Its vertex set is the disjoint union
$$
\bigcup_{n \geq 0} G(n)^{\wedge}
$$
where the set $G(n)^{\wedge}$ is called the $n^{\text{th}}$ level of the diagram. 
They denote by $\varnothing$ the unique element of $G(0)^{\wedge}$.

Writing $S^\lambda$ for the $G(n)$--module that corresponds to an irreducible representation $\lambda \in G(n)^{\wedge}$, they say that two vertices $\mu \in G(n-1)^{\wedge}$ and $\lambda \in G(n)^{\wedge}$ are joined by $k$ directed edges (from $\lambda$ to $\mu$) if
$$
\text{dim Hom}_{G(n-1)}(S^\mu, S^\lambda \downarrow_{G(n-1)}^{G(n)})
= k
$$
that is, if the multiplicity of $\mu$ in the restriction of $\lambda$ from $G(n)$ to $G(n-1)$ is $k$.

They write 
$$
\mu \leftarrow \lambda
$$ 
if $\mu$ and $\lambda$ are connected by at least one edge in the Bratelli diagram, with corresponding notation
$$
S^\mu \subset S^\lambda
$$
for their respective modules.

They show that for the symmetric groups $G(n) = S_n$, the number of edges between any two vertices in adjacent levels of the Bratelli diagram is either $0$ or $1$.
When any chain of finite groups (\ref{chaingroups}) has this property, they say that the \textit{multiplicities are simple} or that the \textit{branching is simple}.

Focusing now on the symmetric group $G(n) = S_n$, since the branching is simple, they obtain a decomposition
$$
S^\lambda \downarrow_{S_{n-1}}^{S_{n}} = \bigoplus_{\substack{\mu \in S_{n-1}^{\wedge} \\ \mu \leftarrow \lambda}} S^\mu
$$
into a direct sum of irreducible $S_{n-1}$ modules. 
This is the restriction version of the branching rule for $S_n$.

Consequently, by using induction on $n$, they obtain a decomposition of $S^\lambda$ into irreducible $S_0$--modules
$$
S^\lambda \downarrow_{S_{0}}^{S_{n}} = \bigoplus_{T} S_T
$$
where the direct sum is over all possible paths of irreducible representations
$$
T = (\lambda_0 \leftarrow \lambda_1 \leftarrow \lambda_2 \leftarrow \dots \leftarrow \lambda_{n-1} \leftarrow \lambda_n = \lambda)
$$
with $\lambda_i \in S_i^{\wedge}$. Note that the $S_T$ are one--dimensional subspaces.

As there exists an $S_n$--invariant inner product for $S^\lambda$, they can choose a unit vector $v_T$ in each $S_T$ with respect to this inner product.
Their union $\{v_T\}_T$ forms a basis of $S^\lambda$, which is called the Gelfand--Tsetlin basis, or GZ--basis, of $S^\lambda$.

Hence, by starting with the chain of symmetric groups 
$$
\{1\} = S_0 \subset S_1 \subset S_2 \subset \dots \subset S_n \subset \dots
$$
and forming the Bratelli diagram,
%relating (between adjacent levels) each group's equivalence classes of irreducibles to one another,
Okounkov and Vershik show that each path
$$
T = (\lambda_0 \leftarrow \lambda_1 \leftarrow \lambda_2 \leftarrow \dots \leftarrow \lambda_{n-1} \leftarrow \lambda_n = \lambda)
$$
through this diagram bijectively corresponds to a chain of irreducible modules 
$$
\varnothing \subset S^{\lambda_1} \subset S^{\lambda_2} \subset \dots \subset S^{\lambda_{n-1}} \subset S^{\lambda_n} = S^\lambda
$$
which corresponds to a basis vector $v_T$ of $S^\lambda$ (that is unique up to a root of unity), all as a direct consequence of the branching being simple.

At this point there has been no mention of partitions or Young tableaux whatsoever in their formulation, yet they obtain many important results in the theory without them.
In fact, Okounkov and Vershik suggest that, given how the Young formulation is developed, the correspondence between partitions and irreducible representations is unnatural, whereas they believe that their approach is more natural and directly leads to the same results. 
They go on to introduce partitions and Young tableaux rather late in their exposition, after they have introduced the Gelfand--Tsetlin algebra $GZ(n)$, the Young--Jucys--Murphy elements which generate it, and some sets which result from this algebra and the GZ-basis formed above.
See \cite{tolli}, \cite{okounkov-english} for more details.

\begin{figure}[ht]
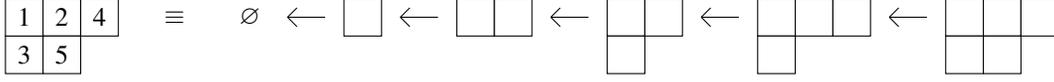

  \ctikzfig{youngpath}
  \caption{An example showing how a standard Young tableau of shape $(3, 2) \vdash 5$ corresponds bijectively to a path through the Young graph.}
  \label{youngpath}
\end{figure}

For our purposes, the only other results that we need from their formulation are that they prove that the Bratelli diagram for $S_n$ is precisely the Young graph; that each irreducible representation $\lambda_n \in S_n^{\wedge}$ is in bijective correspondence with a partition of $n$; and finally that each path (starting with $\lambda \in S_n^{\wedge}$) through the Young graph bijectively corresponds to a standard Young tableau with $n$ boxes.
See Figure \ref{youngpath} for an example of the latter result.
We will use and develop further all of these ideas to achieve a better understanding of the Schur--Weyl basis of $(\mathbb{C}^d)^{\otimes n}$.
%They first call the union of the GZ-bases over all irreducibles $\lambda \in S_n^{\wedge}$ the Young basis, and show (via introduction of the Gelfand--Tsetlin algebra $GZ(n)$) that it is a common eigenbasis of the Young--Jucys--Murphy elements (which they prove generate $GZ(n)$).

% Young shape, partitions, tableaux, paths

\subsection{A Short Introduction to the Representation Theory of $SU(d)$}

The representation theory of $SU(d)$ has a rich history.
%which we won't be able to cover here. 
Interested readers can see \cite{goodman} for more details.
Our interest here lies in particular irreducible representations of $SU(d)$ discovered by Weyl \cite{weyl}, 
where exactly one exists (up to equivalence) for each partition $\lambda \vdash n$, where $n$ is any non-negative integer, such that $\lambda$ has at most $d$ non-zero parts.
As each such irreducible representation corresponds bijectively to its partition $\lambda$, we can denote them by $W^\lambda$.
To describe its basis, we need the following definitions.

\begin{figure}[ht]
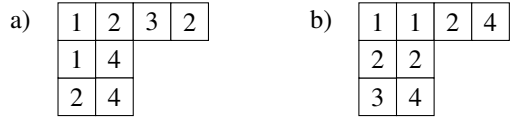

  \ctikzfig{weyltableaux}
	\caption{a) A Weyl tableau for the partition $(4, 2, 2, 0) \vdash 8$ with content $(2,3,1,2)$. It is not standard.
	   b) A standard Weyl tableau for the partition $(4, 2, 2, 0) \vdash 8$ with content $(2,3,1,2)$.}
  \label{weyltableaux}
\end{figure}

For a given $d$, a Weyl tableau of shape $\lambda \vdash n$ consisting of at most $d$ rows with content $(\mu_1, \mu_2, \dots, \mu_d)$, where $\mu_i \geq 0$ for all $i$ and $\sum_{i=1}^{d} \mu_i = n$, is defined to be a Young frame of shape $\lambda$ filled out with entries from the alphabet $\{1, \dots, d\}$ -- hereafter written as $[d]$ -- such that $\mu_i$ boxes have entry $i$, for each $i = 1 \rightarrow d$.
%a filling of a Young frame of shape $\lambda$ such that $\mu_i$ boxes have entry $i$, for each $i = 1 \rightarrow d$.
%Here the entries are said to be taken from the alphabet $\{1, \dots, d\}$, which is hereafter written as $[d]$. 
Such a tableau is said to be standard if the entries in the rows are weakly increasing (from left to right) and the entries in the columns are strictly increasing (from top to bottom).
Figure \ref{weyltableaux}a) shows an example of a Weyl tableau of shape $(4, 2, 2, 0) \vdash 8$ with content $(2,3,1,2)$ that is not standard.
Figure \ref{weyltableaux}b) shows an example of a standard Weyl tableau of shape $(4, 2, 2, 0) \vdash 8$ with content $(2,3,1,2)$.
We note here that when we refer to Weyl tableaux of shape $\lambda$ consisting of at most $d$ rows with entries from the alphabet $[d]$ in the following, it means that we are considering any Weyl tableau of shape $\lambda$ consisting of at most $d$ rows with any allowable $d$-length content.

An equivalent way of representing standard Weyl tableaux of some shape $\lambda \vdash n$ consisting of at most $d$ rows with entries from the alphabet $[d]$ is through so-called Gelfand--Tsetlin patterns.

A Gelfand--Tsetlin pattern is an inverted triangle of numbers 
\begin{equation} \label{GTpattern}
	\begin{pmatrix}
		m_{1, d} & \phantom{=} & m_{2,d} & \phantom{=} & \dots & \phantom{=} & m_{d-1,d} & \phantom{=} & m_{d,d} \\
		\phantom{=} & m_{1, d-1} & \phantom{=} & \phantom{=} & \dots & \phantom{=} & \phantom{=} & m_{d-1,d-1} & \phantom{=} \\
		\phantom{=} & \phantom{=} & \ddots & \phantom{=} & \phantom{=} & \phantom{=} & \iddots \phantom{=} & \phantom{=} \\
		\phantom{=} & \phantom{=} & \phantom{=} & m_{1,2} & \phantom{=} & m_{2,2} & \phantom{=} & \phantom{=} & \phantom{=} \\
		\phantom{=} & \phantom{=} & \phantom{=} & \phantom{=} &  m_{1,1} & \phantom{=} & \phantom{=} & \phantom{=} & \phantom{=}
	\end{pmatrix}
\end{equation}
consisting of some $d$ levels where the numbers in the pattern satisfy the so-called in-betweeness condition
\begin{equation}
	m_{i,j} \geq m_{i, j-1} \geq m_{i+1, j}
	\quad\quad\quad
	1 \leq i < j \leq d
\end{equation}
We denote by $[m]_i$ the numbers appearing of the $i^{\text{th}}$ row of the pattern (\ref{GTpattern}), namely $\{m_{1, i}, \dots, m_{i,i}\}$, and denote by $(m)_i$ the subpattern consisting of the first $i$ rows of the pattern (\ref{GTpattern}). 

The top row in the pattern, $[m]_d$, gives the number of boxes in each row of the standard Weyl tableau to which the pattern corresponds, and so it corresponds to some partition $\lambda$ of some non-negative integer $n$.

We can convert a standard Weyl tableau of shape $\lambda \vdash n$ of at most $d$ rows with entries from the alphabet $[d]$ to a Gelfand--Tsetlin pattern with $d$ levels by recognising that the number of boxes in row $i$ of the tableau with the entry $j$ is equal to $m_{i,j} - m_{i, j-1}$. 
Hence we can use this to fill in the $i^{\text{th}}$ diagonal of the pattern by starting from $j=i$ and iterating through to $j=d$ inclusive. Clearly then $[m]_d = \lambda$.

Conversely, to convert a Gelfand--Tsetlin pattern with $d$ levels to a standard Weyl tableau of shape $[m]_d \vdash n$, we start with $[m]_1$ and place $m_{1,1}$ boxes with entry $1$ in the first row (of what will be our tableau).
Then we take $[m]_2$, and add $m_{1,2} - m_{1,1}$ boxes with entry $2$ to the first row, and $m_{2,2}$ boxes with entry $2$ to the second row.
Iterating through, we take $[m]_j$, and add $m_{1,j} - m_{1, j-1}$ boxes with entry $j$ to the first row, $m_{2,j} - m_{2, j-1}$ boxes with entry $j$ to the second row, and so on, adding $m_{j,j} - m_{j,j-1}$ boxes with entry $j$ to the $j^{\text{th}}$ row.
This will result in the desired standard Weyl tableau.

%\textbf{Refer now to how these appear in the irreps of SU(d) - check Louck?}
%\textbf{Something like this?}
%The representation theory of $SU(d)$ has a rich history which we won't be able to cover here. See \textbf{INSERT REF} for more details.
%Our interest lies in particular irreducibles of $SU(d)$ (the Sym ones?), with exactly one (up to equivalence) existing for each partition $\lambda \vdash n$, where $n$ is any non-negative integer, such that $\lambda$ has at most $d$ parts.
%Hence each such irreducible corresponds bijectively to its partition $\lambda$, and so we denote these by $W^\lambda$.

In fact, these patterns were introduced by Gelfand and Tsetlin precisely for the purpose of labelling a basis of $W^\lambda$, where $\lambda$ is as described above.
The basis of $W^\lambda$ that we are interested in is in bijective correspondence with the set of all $d$-level Gelfand--Tsetlin patterns with the top level $[m]_d$ equal to $\lambda$, or equivalently, with the set of
all standard Weyl tableaux of shape $\lambda$ with entries from the alphabet $[d]$.

We will see this basis appearing in the next section on the Schur--Weyl duality.

\subsection{The Schur--Weyl Duality and the Schur--Weyl Basis} \label{schurbasis}

At its heart, the Schur--Weyl duality is a theorem in representation theory which describes a relationship between the irreducible representations of $S_n$ and the irreducible representations of $SU(d)$ appearing in the decomposition of $(\mathbb{C}^d)^{\otimes n}$ when considered as a representation of each group.

Recall that $\mathbb{C}^d$ has a basis
$
\{\ket{k} \mid k \in [n]\}
$
which is called the computational basis.

Hence $(\mathbb{C}^d)^{\otimes n}$ has as its computational basis
$$
\{ \ket{v_1} \otimes \ket{v_2} \otimes \dots \otimes \ket{v_n} \mid v_i \in [d] \text{ for all } i = 1 \rightarrow n \}
$$
We know that $(\mathbb{C}^d)^{\otimes n}$ is a representation of the symmetric group $S_n$ -- written $\rho: S_n \rightarrow GL((\mathbb{C}^d)^{\otimes n})$ -- under
\begin{equation}
	\rho(\sigma)(\ket{v_1} \otimes \ket{v_2} \otimes \dots \otimes \ket{v_n})
	=
	\ket{v_{\sigma^{-1}(1)}} \otimes \ket{v_{\sigma^{-1}(2)}} \otimes \dots \otimes \ket{v_{\sigma^{-1}(n)}}
\end{equation}
where $\sigma \in S_n$, $\ket{v_i}$ is a computational basis state of $\mathbb{C}^d$, and $\sigma^{-1}(i)$ is the label resulting from the action of $\sigma^{-1}$ on the index $i$.

Similarly, $(\mathbb{C}^d)^{\otimes n}$ is also a representation of the special unitary group $SU(d)$ -- written $\tau: SU(d) \rightarrow GL((\mathbb{C}^d)^{\otimes n})$ -- under
\begin{equation}
	\tau(U)(\ket{v_1} \otimes \ket{v_2} \otimes \dots \otimes \ket{v_n})
	=
	U\ket{v_1} \otimes U\ket{v_2} \otimes \dots \otimes U\ket{v_n}
\end{equation}
where $U \in SU(d)$ and, again, $\ket{v_i}$ is a computational basis state of $\mathbb{C}^d$. Clearly $\tau(U) = U^{\otimes n}$.

As $\rho$ and $\tau$ are both representations of groups that admit a decomposition into irreducible representations, we get that
\begin{equation} \label{symmetricdecomp}
	(\mathbb{C}^d)^{\otimes n} 
	\cong
	\bigoplus_{\lambda} \mathbb{C}^{m_\lambda} \otimes S^{\lambda}
\end{equation}
as a representation of $S_n$ and
\begin{equation}
	(\mathbb{C}^d)^{\otimes n} 
	\cong
	\bigoplus_{\mu} W^{\mu} \otimes \mathbb{C}^{n_\mu}
\end{equation}
as a representation of $SU(d)$.
Here, $\lambda$ and $\mu$ run over the set of partitions of $n$ with at most $d$ parts.
Also, $m_\lambda$ is the multiplicity of the irreducible $S^{\lambda}$ in its corresponding decomposition, and, likewise, $n_\mu$ is the multiplicity of the irreducible $W^{\mu}$ in its corresponding decomposition.

%We obtain the duality from looking further into the structure of these representations.
%We can see that for any $\sigma \in S_n$ and $U \in SU(d)$, we have that
%\begin{equation}
%	\rho(\sigma)\tau(U) 
%	= 
%	\tau(U)\rho(\sigma)
%\end{equation}
%that is, that the representations commute.

%This implies that the irreducible representations of $\rho$ must act on the multiplicity labels of $\tau$ and vice versa, via Schur's Lemma. \textbf{WHY? - see Goodman and Wallach for more details}.

%Moreover, we can say that the group algebras that they generate centralise each other \textbf{INSERT MORE HERE}.

However, by looking further into the structure of these representations, we can, in fact, say that $(\mathbb{C}^d)^{\otimes n}$ is a representation of the direct product group $SU(d) \times S_n$, and that its decomposition into irreducible representations is given by
\begin{equation} \label{schurweylduality}
	(\mathbb{C}^d)^{\otimes n} 
	\cong
	\bigoplus_{\lambda} W^{\lambda} \otimes S^{\lambda}
\end{equation}
where $m_\lambda = \text{dim } W^\lambda$ and $n_\lambda = \text{dim } S^\lambda$, and where $\lambda$ runs over the set of partitions of $n$ with at most $d$ parts.

The isomorphism given in (\ref{schurweylduality}) is known as the Schur--Weyl duality. See \cite{goodman} for more details.

From the duality we obtain the so-called Schur--Weyl basis of $(\mathbb{C}^d)^{\otimes n}$, which consists of a set of triplets of the form
$$
\ket{\lambda t y}
$$
where $\lambda$ is a partition of $n$ with at most $d$ parts, 
$t$ is a standard Weyl tableau of shape $\lambda$ with entries from the alphabet $[d]$ 
and $y$ is a standard Young tableau of shape $\lambda$. 

This basis forms the focus of our attention for the rest of this paper.

\section{The Schur--Weyl--Young Graph and the Schur--Weyl Branching Rule}
\label{mainresults}

%We have come to the main section of the paper which describes 
We now present our main contributions.
Specifically, we present a new way of understanding how the Schur--Weyl duality can be used to perform the Quantum Schur Transform.
We achieve our results by taking inspiration from the Okounkov--Vershik formulation for the representation theory of the symmetric groups.
%taking inspiration from the Okounkov--Vershik approach to the representation theory of the symmetric groups.

%We will use the ideas given here to present, in Section \ref{schurtransform}, a new method for constructing the Quantum Schur Transform --
%a unitary transformation mapping the computational basis of $(\mathbb{C}^d)^{\otimes n}$ to the Schur--Weyl basis of $(\mathbb{C}^d)^{\otimes n}$ --
%in \textbf{WHAT TIME} time.

%Over the following sections, we describe a remarkable result which relates the computational basis of $(\mathbb{C}^d)^{\otimes n}$ to the Schur-Weyl basis of $(\mathbb{C}^d)^{\otimes n}$, and leads to a new way of understanding and constructing the Quantum Schur Transform in \textbf{WHAT TIME} time.

%We recall that the Okounkov--Vershik formulation takes as its starting point a chain of groups $G(n) = S_n$ and from this obtains chains of irreducible representations that correspond to paths through the Young diagram, where, furthermore, each path is in bijective correspondence with some standard Young tableau.

%Consider now instead the following chain of groups
Consider the following chain of direct product groups
\begin{equation} \label{schurweylchain}
SU(d) \times S_0 \subset
SU(d) \times S_1 \subset SU(d) \times S_2 \subset \dots \subset SU(d) \times S_n 
\end{equation}
where we have fixed $d$ and (in the following) have stopped the chain at level $n$ for some $n$. 

We have, corresponding to this chain of groups, a chain of representations 
\begin{equation} \label{dnchain}
\varnothing
\subset 
(\mathbb{C}^d)^{\otimes 1}
\subset 
(\mathbb{C}^d)^{\otimes 2}
\subset \dots 
\subset 
(\mathbb{C}^d)^{\otimes n}
\end{equation}
as discussed in Section \ref{background}.

However, in following the Okounkov--Vershik formulation as inspiration, we are particularly interested in finding chains of irreducible representations for the chain of groups (\ref{schurweylchain}).

We can use the chain of representations given in (\ref{dnchain}) to construct chains of irreducible representations in the following way.

By the Okounkov--Vershik formulation, for every partition $\lambda$ of $n$ consisting of at most $d$ non-zero parts, we can choose a path through the Young graph
\begin{equation} \label{youngpathlambda}
T = (\lambda_0 \leftarrow \lambda_1 \leftarrow \lambda_2 \leftarrow \dots \leftarrow \lambda_{n-1} \leftarrow \lambda_n = \lambda)
\end{equation}
where each $\lambda_i$ is a partition of $i$ with at most $d$ non-zero parts.
In fact, for each such $\lambda$, we get $\text{dim } S^{\lambda}$ many paths, which are in bijective correspondence with the standard Young tableaux of shape $\lambda$.

The key point is that, for each partition $\lambda_i$ appearing in the path $T$, $S^{\lambda_i}$ appears in the irreducible decomposition of $(\mathbb{C}^d)^{\otimes i}$ when viewed as a representation of $S_i$.
This follows by the isomorphism given in (\ref{symmetricdecomp}).
(This isomorphism is also why we only consider partitions of $n$ with at most $d$ non-zero parts in the above.)

We also know, by the same formulation, that the path given in (\ref{youngpathlambda}) corresponds to a chain 
$$
\varnothing \subset
S^{\lambda_1} \subset S^{\lambda_2} \subset \dots \subset S^{\lambda_{n-1}} \subset S^{\lambda_n} 
$$
of irreducible representations for the chain of symmetric groups
$$
\{1\} = S_0 \subset S_1 \subset S_2 \subset \dots \subset S_n 
$$
Hence, by the Schur--Weyl duality (\ref{schurweylduality}), we can tensor on the accompanying irreducible representation of $SU(d)$ (that is determined by the duality) to each irreducible representation in this chain, leading to the chain of irreducible representations
\begin{equation} \label{schurweylirrepchain}
\varnothing \subset
W^{\lambda_1} \otimes S^{\lambda_1}
\subset 
W^{\lambda_2} \otimes S^{\lambda_2}
\subset \dots 
\subset 
W^{\lambda_n} \otimes S^{\lambda_n}
\end{equation}
corresponding to the chain of direct product groups given in (\ref{schurweylchain}).
Each $W^{\lambda_i} \otimes S^{\lambda_i}$ appears in the decomposition of $(\mathbb{C}^d)^{\otimes i}$ into irreducible representations when viewed as a representation of $SU(d) \times S_i$.

This leads to the discovery of a new multigraph, which, to the best of our knowledge, has not been described previously in the literature.
We modify the vertex and edge sets of the Young graph for the symmetric groups in the following way.
Firstly, the vertex set:
for a given $d$, remove all Young frames that have more than $d$ rows. 
Then, for each remaining Young frame, calculate all of the possible standard Weyl tableaux of the same shape using elements from the alphabet $[d]$, and replace the Young frame by the set of all such Weyl tableaux. 
(Note that for $d=2$ we will use the alphabet $\{0, 1\}$ and not $\{1, 2\}$ to align our notation with that for the computational basis of a qubit.)
Then, for the edge set:
for each (single) edge that appears between Young frames in the Young graph, we replace it with an edge between an element from each of the sets that have replaced them, that is, between standard Weyl tableaux, if the following holds.
%We replace each (single) path appearing between Young frames in the Young graph with a path between an element of each of the sets that have replaced them in the following way:
Suppose that $\mu \vdash n-1$ and $\lambda \vdash n$ are partitions that had an edge between them in the Young graph.
Let $t^{\mu}$ be a standard Weyl tableau in the set replacing the Young frame of shape $\mu$, and let $t^{\lambda}$ be a standard Weyl tableau in the set replacing the Young frame of shape $\lambda$.
Then an edge exists between these tableaux if and only if $t^{\mu}$ can be obtained from $t^{\lambda}$ by removing an element of the alphabet $[d]$ from it; equivalently, if $t^{\lambda}$ can be obtained from $t^{\mu}$ by adding an element of the alphabet $[d]$ to it.
We call such an edge a \textit{transition} between standard Weyl tableaux.
Clearly, between the sets appearing in this new graph, there may be multiple edges.

We will call this multigraph the Schur--Weyl--Young $SU(d)-S_n$ graph (or the Schur--Weyl--Young graph when $n$ and $d$ are clearly understood). 
An example for $d=2$, up to and including $n = 3$, is given in Figure \ref{weyldiagramd2}.

We will use the Schur--Weyl--Young graph to calculate, 
for pairs of standard Weyl tableaux that have an edge between them, 
%in the next section.
the transition amplitude (a term whose definition is delayed to Section \ref{patternrulessection}).

\begin{figure}[ht]
  \ctikzfig{weyldiagramd2}
  \caption{The Schur--Weyl--Young $SU(2)-S_n$ graph up to and including $n = 3$. Note that in our presentation of the graph, the rows of a given level have been repeated to show clearly all of the possible transitions between the standard Weyl tableaux.}
  \label{weyldiagramd2}
\end{figure}

What seems rather remarkable is that, as a direct consequence of starting with paths through the Young graph,
we can construct a branching rule for the Schur--Weyl states.
%what we will call the Schur-Weyl branching rule. 

Indeed, we have shown that for the chain of direct product groups (\ref{schurweylchain})
$$
SU(d) \times S_0 \subset
SU(d) \times S_1 \subset SU(d) \times S_2 \subset \dots \subset SU(d) \times S_n 
$$
we can form a chain of irreducible representations (\ref{schurweylirrepchain})
$$
\varnothing
\subset 
W^{\lambda_1} \otimes S^{\lambda_1}
\subset 
W^{\lambda_2} \otimes S^{\lambda_2}
\subset \dots 
\subset 
W^{\lambda_n} \otimes S^{\lambda_n}
$$
corresponding to a path (\ref{youngpathlambda})
%where each $W^{\lambda_i} \otimes S^{\lambda_i}$ appears in the decomposition of $(\mathbb{C}^d)^{\otimes i}$ into irreducibles as a representation of $SU(d) \times S_i$, 
%which corresponds to a path 
$$
T = (\lambda_0 \leftarrow \lambda_1 \leftarrow \lambda_2 \leftarrow \dots \leftarrow \lambda_{n-1} \leftarrow \lambda_n)
$$
through the Young graph, where each $\lambda_i$ is a partition of $i$ with at most $d$ non-zero parts.

We noted that each $W^{\lambda_i} \otimes S^{\lambda_i}$ appears in the decomposition of $(\mathbb{C}^d)^{\otimes i}$ into irreducible representations when viewed as a representation of $SU(d) \times S_i$.
%(Equivalently, it corresponds to a standard Young tableaux of shape $\lambda_n \vdash n$.)

Consequently, given the chain (\ref{dnchain})
$$
\varnothing
\subset 
(\mathbb{C}^d)^{\otimes 1}
\subset 
(\mathbb{C}^d)^{\otimes 2}
\subset \dots 
\subset 
(\mathbb{C}^d)^{\otimes n}
$$
we see that, as
$$
(\mathbb{C}^d)^{\otimes i-1} \otimes \mathbb{C}^d
=
(\mathbb{C}^d)^{\otimes i}
$$
we must have, from (\ref{schurweylirrepchain}), that
\begin{equation} \label{swbranching}
(W^{\lambda_{i-1}} \otimes S^{\lambda_{i-1}}) \otimes \mathbb{C}^d
=
W^{\lambda_{i}} \otimes S^{\lambda_{i}} 
\end{equation}

%\textbf{Check that this gives me what I really want. By this I mean I really need to discuss why this works, and that this equation actually gives me 2 versions of the branching rule, i.e going left--to--right over the equation and also going right--to--left}

%\textbf{Get the notation consisent between i and n}.

In obtaining the equality given in (\ref{swbranching}), we have actually found a branching rule for the Schur--Weyl states that is intimately connected to the Schur--Weyl--Young diagram.
It comes in two versions that are inverse operations of one another, depending on whether we start on the left-hand side or the right-hand side of the equality given in (\ref{swbranching}).

The left-to-right version takes a pair of quantum registers, the first a Schur--Weyl basis state of $(\mathbb{C}^d)^{\otimes i-1}$, the second a computational basis state of $\mathbb{C}^d$, and gives an equivalent expression in terms of the Schur--Weyl basis of $(\mathbb{C}^d)^{\otimes i}$.

The right-to-left version is the inverse of this: it takes a Schur--Weyl basis state of $(\mathbb{C}^d)^{\otimes i}$ and gives its equivalent expression in terms of a superposition of pairs of quantum registers: in each pair, the first register is a Schur--Weyl basis state of $(\mathbb{C}^d)^{\otimes i-1}$ and the second register is a computational basis state of $\mathbb{C}^d$.

Together, these versions make up what we shall call, from now on, the Schur--Weyl branching rule.

%\textbf{Potentially edit this/put after we show how the branching rule works}
%because, starting from a computational basis state of $(\mathbb{C}^d)^{\otimes n}$ and iteratively applying the left--to--right version on each qudit (i.e starting from the left--most qudit) will give the same state expressed in terms of the Schur--Weyl basis of $(\mathbb{C}^d)^{\otimes n}$ and, similarly,
%starting from a Schur--Weyl basis state of $(\mathbb{C}^d)^{\otimes n}$ and iteratively applying the right--to--left version will give the same state expressed in terms of the computational basis of $(\mathbb{C}^d)^{\otimes n}$.

%We discuss this further in Section \ref{schurtransform}.

\subsection{How does the Schur--Weyl branching rule work?}

\textbf{Left-to-right version}

Take as input a pair of quantum registers, the first a Schur--Weyl basis state of $(\mathbb{C}^d)^{\otimes i-1}$, the second a computational basis state of $\mathbb{C}^d$.

Therefore, the input is of the form $\ket{\mu t y}\ket{k}$, where $\mu \vdash i-1$ consists of at most $d$ non-zero parts, $t$ is a standard Weyl tableau of shape $\mu$ with entries from the alphabet $[d]$, $y$ is a standard Young tableau of shape $\mu$, and $k \in [d]$.

Consider now each possible addition of a box with entry $i$ to the standard Young tableau $y$ such that a new Young tableau $y^{*}$ of some shape $\lambda \vdash i$ consisting of at most $d$ rows is formed.
Clearly, $y^{*}$ will also be standard.
Each possibility $y^{*}$ corresponds to elongating the path of partitions $T$ that is determined by $y$ with $\lambda$, as suggested by equation (\ref{swbranching}), or, equivalently, to the induction version of the symmetric group branching rule.
%This corresponds to the induction version of the symmetric group branching rule, where, equivalently, each possibility corresponds to elongating the path of partitions $T$ that is determined by $y$ with $\lambda$.

Then, for each one of these new standard Young tableau $y^{*}$, find all possible edges in the Schur--Weyl--Young graph 
between $t$, the standard Weyl tableau of shape $\mu$ given in the first register, and $t^{*}$, a standard Weyl tableau of the new shape $\lambda$, where the element $k$ (given in the second register) has been added into $t$ to give $t^{*}$.
%(such that it is standard of shape $\lambda$).

Each new Young tableau and each such edge between the standard Weyl tableaux (as described above) results in a Schur--Weyl state $\ket{\lambda t^{*} y^{*}}$ of $(\mathbb{C}^d)^{\otimes i}$. 

Form as output a superposition over all such states, with the amplitude for each state appearing in the superposition given by the transition amplitude between $t$ and $t^{*}$.
%Suppose that the second register is $\ket{k}$, where $k \in [d]$.
%Form all possible insertions of $k$ into the $i-1$--size standard Weyl tableau of the first register such that we obtain a standard $i$--size Weyl tableau, updating its accompanying Young tableau accordingly. \textbf{HOW ARE THE INSERTIONS PERFORMED?}
%This will give Schur--Weyl basis states of $(\mathbb{C}^d)^{\otimes n+1}$.
%Form a superposition of these new states, with each amplitude given by the transition amplitude between the input Weyl tableau and the Weyl tableau of the newly formed state.
%This output is an equivalent expression for the input in terms of the Schur--Weyl basis of $(\mathbb{C}^d)^{\otimes n+1}$.

\textbf{Right-to-left version}

Take as input a Schur--Weyl basis state of $(\mathbb{C}^d)^{\otimes i}$, that is, $\ket{\lambda t y}$, where $\lambda \vdash i$ consists of at most $d$ non-zero parts, $t$ is a standard Weyl tableau of shape $\lambda$ with entries from the alphabet $[d]$, and $y$ is a standard Young tableau of shape $\lambda$.

Let $y^{*}$ be the standard Young tableau of shape $\mu \vdash i-1$, where $\mu$ is of the same shape as $y$ but with the box with entry $i$ removed.
Note that we have used equation (\ref{swbranching}) here, since it tells us that the standard Young tableau appearing on the left-hand side is pre-determined by $y$ (as $y$ corresponds bijectively to a path of partitions $T$ through the Young graph).

Now, for each $k \in [d]$, find all possible edges
in the Schur--Weyl--Young graph
between $t$, the standard Weyl tableau of shape $\lambda$ given as input, and $t^{*}$, a standard Weyl tableau of the new shape $\mu$ with a single $k$ removed.
%, of the same shape as the Young tableau with the box labelled by $n$ removed.

Each such edge between the standard Weyl tableaux together with the new Young tableau $y^{*}$ results in a Schur--Weyl basis state of the form $\ket{\mu t^{*} y^{*}}$ of $(\mathbb{C}^d)^{\otimes i-1}$.

This state becomes the first register in a pair of quantum registers. Let the second register be $\ket{k}$, where $k$ is the entry removed from $t$ to form $t^{*}$.
%For each such state, let $\ket{k}$ (where $k$ is the entry removed from the input Weyl tableau $t$ to form $t^{*}$) be the second register for each such state.

Form as output a superposition of these new pairs of quantum registers, 
with the amplitude for each state appearing in the superposition given by the transition amplitude between $t^{*}$ and $t$.
%with each amplitude given by the transition amplitude between $t$ and $t^{*}$.
%the Weyl tableau given in the first register of the newly formed pair and the input Weyl tableau.
%This is the desired output, as described above.

\subsubsection{Remarks} 

To end this section, we make clear how these new ideas have been inspired by the Okounkov--Vershik formulation for the representation theory of the symmetric groups.
%descriptions of the two versions of the Schur--Weyl Branching Rule, we see how intimately involved the previously described ideas are.

In this formulation, 
%Okounkov-Vershik formulation for the representation theory of the symmetric groups, 
we saw that a path beginning at some $\lambda \vdash n$ through the Young graph corresponds bijectively to a standard Young tableau of shape $\lambda$,
%a standard Young tableau of shape $\lambda \vdash n$ corresponds bijectively to a path through the Young graph, 
which corresponds bijectively to a chain of irreducible representations relating to the chain of groups $G(n) = S_n$.
Also, the Young graph corresponds to a branching rule for irreducible representations of $S_n$.

In the above, we have shown that, for a given $d$, a pair of tableaux -- the first a standard Weyl tableau of shape $\lambda \vdash n$ consisting of at most $d$ rows with entries from the alphabet $[d]$, the second a standard Young tableau of the same shape $\lambda$ -- corresponds to a multipath through the (newly formed) Schur--Weyl--Young graph, which corresponds to a chain of irreducible representations coming from the Schur--Weyl duality relating to the chain of groups $G(n) = SU(d) \times S_n$.
We did this by starting with paths through the Young graph -- which correspond bijectively to standard Young tableaux~-- and modifying them to pair them with standard Weyl tableaux.
As a result, we showed that the Schur--Weyl--Young graph corresponds to a branching rule for the Schur--Weyl basis states of $(\mathbb{C}^d)^{\otimes n}$, a representation of the direct product group $SU(d) \times S_n$.

The Schur--Weyl branching rule is important, because, as we will discuss further in Section \ref{schurtransform}, starting from a computational basis state of $(\mathbb{C}^d)^{\otimes n}$ and iteratively applying the left-to-right version on each qudit -- beginning at the left-most qudit -- gives the same state expressed in terms of the Schur--Weyl basis of $(\mathbb{C}^d)^{\otimes n}$ and, similarly,
starting from a Schur--Weyl basis state of $(\mathbb{C}^d)^{\otimes n}$ and iteratively applying the right-to-left version gives the same state expressed in terms of the computational basis of $(\mathbb{C}^d)^{\otimes n}$.

We delay giving examples of these versions of the Schur--Weyl branching rule until Section \ref{examplesswbranchingrule}, after
we have described how to calculate the transition amplitudes between standard Weyl tableaux, which we do in the next section.

\section{Simple Pattern Rules for the Calculation of Transition Amplitudes when $d = 2$} \label{patternrulessection}

As we have seen, an element of a vertex in the Schur--Weyl--Young graph at level $n$ is a standard Weyl tableau $t$ of some shape $\lambda \vdash n$ consisting of at most $d$ rows with entries from the alphabet $[d]$.
This tableau has a number of connections to standard Weyl tableaux $t^{*}$ in the $n+1^{\text{st}}$ level, where each connection is given by an edge in the graph.
Each edge therefore describes a possible transition of $t$ to one such $t^{*}$ in the next level.
Hence, in moving up a level (by the addition of an entry from the alphabet $[d]$ into $t$ such that we obtain a standard Weyl tableau of some shape with $n+1$ entries and at most $d$ rows), the new Weyl tableau must be one of the $t^{*}$.
Consequently, each edge must come with a so-called transition amplitude whose square gives the probability that we will obtain the new standard Weyl tableau $t^{*}$ out of all such possibilities.

%An edge between two vertices of the Schur--Weyl--Young graph describes a transition between two Weyl tableaux with elements from the alphabet $[d]$.
%(Their shapes are given by the Young frames they inhabit.)
%... \textbf{Discuss paths through this diagram and how that generates transition amplitudes}.

Louck \cite{louck} has shown (independently of the existence of the Schur--Weyl--Young graph) that the transition amplitude, for general $d$, between two standard Weyl tableaux that are in adjacent levels of the Schur--Weyl--Young diagram and are connected by an edge (denoted by $\longleftrightarrow$ in the following)
\begin{equation} \label{transition}
	\begin{pmatrix}
		[m]_d \\
		[m]_{d-1} \\
		\vdots \\
		[m]_k \\
		(m)_{k-1}
	\end{pmatrix}
	\longleftrightarrow
	\begin{pmatrix}
		[m]_d + e_d(\tau_d) \\
		[m]_{d-1} +  e_{d-1}(\tau_{d-1}) \\
		\vdots \\
		[m]_k + e_k(\tau_k) \\
		(m)_{k-1}
	\end{pmatrix}
\end{equation}
where each Weyl tableau has been expressed in its equivalent Gelfand--Tsetlin pattern form, is given by the multiplication of two terms, the first of which is 
\begin{equation} \label{louckfirst}
	\prod_{j=k+1}^d \text{sgn}(\tau_{j-1} - \tau_{j})
	\sqrt{
		\left|
		\frac{
			\prod_{i=1, i \neq \tau_{j-1}}^{j-1} (p_{\tau_j, j} - p_{i, j-1})
			\prod_{i=1, i \neq \tau_{j}}^{j} (p_{\tau_{j-1}, j-1} - p_{i, j} + 1)
		}
		{
			\prod_{i=1, i \neq \tau_{j}}^{j} (p_{\tau_{j}, j} - p_{i, j})
			\prod_{i=1, i \neq \tau_{j-1}}^{j-1} (p_{\tau_{j-1}, j-1} - p_{i, j-1} + 1)
		}
		\right|
	}
\end{equation}
and the second of which is
\begin{equation} \label{loucksecond}
	\sqrt{
		\left|
		\frac{
			\prod_{i=1}^{k-1} (p_{\tau_k, k} - p_{i, k-1})
		}
		{
			\prod_{i=1, i \neq \tau_{k}}^{k} (p_{\tau_{k}, k} - p_{i, k})
		}
		\right|
	}
\end{equation}
unless $k=1$, where the second term (\ref{loucksecond}) in the multiplication is defined to be $1$, and if $k=d$, the first term (\ref{louckfirst}) in the multiplication is defined to be $1$.

%Here, $k \in [d]$ is the entry that is being inserted into the first (lower level) Weyl tableau to form the second (higher level) Weyl tableau.
Here, $k \in [d]$ can be thought of either as the entry that is added to the first (lower level) Weyl tableau to form the second (higher level) Weyl tableau, or as the entry that is removed from the second Weyl tableau to form the first Weyl tableau.

%$[m]_i$, for each $i: k \rightarrow d$, is defined to be the $i^{\text{th}}$ row of the lower level Gelfand--Tsetlin pattern.
%In Weyl tableau form, it is the tableau formed from the entries $1 \rightarrow i$ inclusive appearing in the lower level Weyl tableau.

%$(m)_{k-1}$ is the subpattern consisting of the first $k-1$ rows of the lower level Gelfand--Tsetlin pattern.

%We call $p_{i,j}$ the partial hook, and it is defined to be equal to $m_{i,j} + j - i$.
The integer $p_{i,j}$ appearing in (\ref{louckfirst}) and (\ref{loucksecond}) is called the partial hook, and it is defined to be equal to $m_{i,j} + j - i$.

Finally, $e_i(\tau_i)$ is a vector of length $i$, with entry $1$ in the position $\tau_i$, and 0 otherwise. 
$\tau_i$ itself is found by creating Weyl sub--tableaux formed from the entries $1 \rightarrow i$ inclusive appearing in each of the Weyl tableaux involved and then looking at which row in the sub--tableaux has an extra box in it. $\tau_i$ is equal to this row number.

Whilst the above formula is true, it is rather involved, hence using it in practice to calculate the transition amplitude between standard Weyl tableaux is slow and fraught with danger.
For the case $d=2$, we have found a much simpler way of calculating these transition amplitudes that is based entirely on looking at the entries in the standard Weyl tableaux themselves.
In fact, we have reduced the entire calculation to the application of four simple rules for this case.
From now on, we will call these rules the Pattern Rules.

We give the Pattern Rules both in their Weyl tableaux form and in their equivalent Gelfand--Tsetlin form below.
A proof can be found in Appendix \ref{patternproof}.

\subsection{The Pattern Rules for $d=2$} \label{patternrules}

Denote the transition amplitude between the standard Weyl tableaux (or Gelfand--Tsetlin patterns) by $\alpha$.
In the following, the notation $\mu \longleftrightarrow \lambda$ means that the transition is between standard Weyl tableaux of shapes $\mu$ and $\lambda$ respectively.

\textbf{Rule 1} 

\textit{In Weyl tableaux form:} $(n-1, 0) 
	\longleftrightarrow
	%\text{ --- }
%\longrightarrow 
(n,0)$

%Delete all boxes containing a $1$ in both tableaux.

a) If the number of zeroes in each tableaux are the same, then 
%a) If the resulting tableau are the same, then 
\begin{equation}
	\alpha = \sqrt{
		\frac{
			\text{Number of ones in the } (n,0) \text{ tableau}
		}{n}
	}
\end{equation}
b) Else
\begin{equation}
	\alpha = \sqrt{
		\frac{
			\text{Number of zeroes in the } (n,0) \text{ tableau}
		}
		{n}
	}
\end{equation}

\textit{In Gelfand--Tsetlin pattern form:}
$$
	\Bigl(
	\begin{smallmatrix}
		n-1 & \phantom{=} & 0 \\
		\phantom{=} & m_{1,1}^{n-1} & \phantom{=}
	\end{smallmatrix}
	\Bigr)
	\longleftrightarrow
	%\text{ --- }
	%\rightarrow
	\Bigl(
	\begin{smallmatrix}
		n & \phantom{=} & 0 \\
		\phantom{=} & m_{1,1}^{n} & \phantom{=}
	\end{smallmatrix}
	\Bigr)
$$

a) If $m_{1,1}^{n-1} = m_{1,1}^{n}$, then
%the bottom number in each Gelfand--Tsetlin pattern is the same, then 
\begin{equation} \label{rule1a}
	\alpha = \sqrt{
		\frac{
			n - m_{1,1}^{n}
		}{n}
	}
\end{equation}
b) Else
\begin{equation} \label{rule1b}
	\alpha = \sqrt{
		\frac{
			m_{1,1}^{n}
		}{n}
	}
\end{equation}
%where $m_{1,1}$ is the bottom entry in the size $n$ Gelfand--Tsetlin pattern.

\textbf{Rule 2} 

\textit{In Weyl tableaux form:} $(n-1, 0) 
\longleftrightarrow
%\text{ --- }
%\longrightarrow 
(n-1, 1)$

%Delete all boxes containing a $1$ in both tableaux.

a) If the number of zeroes in each tableaux are the same, then 
%a) If the resulting tableau are the same, then 
\begin{equation}
	\alpha = \sqrt{
		\frac{
			\text{Number of zeroes in the } (n,0) \text{ tableau}
		}
		{n}
	}
\end{equation}
b) Else
\begin{equation}
	\alpha = - \sqrt{
		\frac{
			\text{Number of ones in the } (n,0) \text{ tableau}
		}{n}
	}
\end{equation}

\textit{In Gelfand--Tsetlin pattern form:}
$$
	\Bigl(
	\begin{smallmatrix}
		n-1 & \phantom{=} & 0 \\
		\phantom{=} & m_{1,1}^{n-1} & \phantom{=}
	\end{smallmatrix}
	\Bigr)
	\longleftrightarrow
	%\text{ --- }
	%\rightarrow
	\Bigl(
	\begin{smallmatrix}
		n-1 & \phantom{=} & 1 \\
		\phantom{=} & m_{1,1}^{n} & \phantom{=}
	\end{smallmatrix}
	\Bigr)
$$

a) If $m_{1,1}^{n-1} = m_{1,1}^{n}$, then
%a) If the bottom number in each Gelfand--Tsetlin pattern is the same, then 
\begin{equation} \label{rule2a}
	\alpha = \sqrt{
		\frac{
			m_{1,1}^{n}
		}{n}
	}
\end{equation}
b) Else
\begin{equation} \label{rule2b}
	\alpha = - \sqrt{
		\frac{
			n - m_{1,1}^{n}
		}{n}
	}
\end{equation}
%where $m_{1,1}$ is the bottom entry in the size $n$ Gelfand--Tsetlin pattern.

\newpage

\textbf{Rule 3} 

\textit{In Weyl tableaux form:} $(n-k-1, k)
\longleftrightarrow
%\text{ --- }
%\longrightarrow 
(n-k, k)$

%Here the Weyl tableaux have $k$ shared columns.
Here the first $k$ columns are the same in both Weyl tableaux.
Delete all $k$ such columns from both tableaux. 
Then apply Rule 1.

\textit{In Gelfand--Tsetlin pattern form:}
$$
	\Bigl(
	\begin{smallmatrix}
		n-k-1 & \phantom{=} & k \\
		\phantom{=} & m_{1,1}^{n-1} & \phantom{=}
	\end{smallmatrix}
	\Bigr)
	\longleftrightarrow
	%\text{ --- }
	%\rightarrow
	\Bigl(
	\begin{smallmatrix}
		n-k & \phantom{=} & k \\
		\phantom{=} & m_{1,1}^{n} & \phantom{=}
	\end{smallmatrix}
	\Bigr)
$$

Subtract $k$ from every entry in both patterns.
Then apply Rule 1.

\textbf{Rule 4} 

\textit{In Weyl tableaux form:} $(n-k, k-1) 
\longleftrightarrow
%\text{ --- }
%\longrightarrow 
(n-k, k)$

%Here the Weyl tableaux have $k-1$ shared columns.
Here the first $k-1$ columns are the same in both Weyl tableaux.
Delete all $k-1$ such columns from both tableaux. 
Then apply Rule 2.

\textit{In Gelfand--Tsetlin pattern form:}
$$
	\Bigl(
	\begin{smallmatrix}
		n-k & \phantom{=} & k-1 \\
		\phantom{=} & m_{1,1}^{n-1} & \phantom{=}
	\end{smallmatrix}
	\Bigr)
	\longleftrightarrow
	%\text{ --- }
	%\rightarrow
	\Bigl(
	\begin{smallmatrix}
		n-k & \phantom{=} & k \\
		\phantom{=} & m_{1,1}^{n} & \phantom{=}
	\end{smallmatrix}
	\Bigr)
$$

Subtract $k-1$ from every entry in both patterns.
Then apply Rule 2.

\subsection{Examples}

\textbf{Example 1: }
$
\ytableausetup{smalltableaux}
\begin{ytableau}
	0 & 0 \\
	1
\end{ytableau}
\longleftrightarrow
\ytableausetup{smalltableaux}
\begin{ytableau}
	0 & 0 & 1 \\
	1
\end{ytableau}
$

\textit{In Weyl tableaux form:}

Apply rule 3 to the one shared column:
$
\ytableausetup{smalltableaux}
\begin{ytableau}
	0
\end{ytableau}
\longleftrightarrow
\ytableausetup{smalltableaux}
\begin{ytableau}
	0 & 1
\end{ytableau}
$

Now apply rule 1a) to these tableaux: $\alpha = \frac{1}{\sqrt{2}}$.

\textit{In Gelfand--Tsetlin pattern form:}

The transition is
$
	\bigl(
	\begin{smallmatrix}
		2 & \phantom{=} & 1 \\
		\phantom{=} & 2 & \phantom{=}
	\end{smallmatrix}
	\bigr)
	\longleftrightarrow
	\bigl(
	\begin{smallmatrix}
		3 & \phantom{=} & 1 \\
		\phantom{=} & 2 & \phantom{=}
	\end{smallmatrix}
	\bigr)
$. 

Apply rule 3. 
%The smallest value across both patterns is $1$. 
Subtracting $1$ from every entry in both patterns gives
$
	\bigl(
	\begin{smallmatrix}
		1 & \phantom{=} & 0 \\
		\phantom{=} & 1 & \phantom{=}
	\end{smallmatrix}
	\bigr)
	\longleftrightarrow
	\bigl(
	\begin{smallmatrix}
		2 & \phantom{=} & 0 \\
		\phantom{=} & 1 & \phantom{=}
	\end{smallmatrix}
	\bigr)
$.

Now apply rule 1a) to these patterns: 
%as the bottom number in each pattern is the same, we have that 
$\alpha = \frac{1}{\sqrt{2}}$.

\textbf{Example 2: }
$
\ytableausetup{smalltableaux}
\begin{ytableau}
	0 & 0 & 1 \\
	1 & 1
\end{ytableau}
\longleftrightarrow
\ytableausetup{smalltableaux}
\begin{ytableau}
	0 & 0 & 0 \\
	1 & 1 & 1
\end{ytableau}
$

\textit{In Weyl tableaux form:}

Apply rule 4 to the two shared columns:
$
\ytableausetup{smalltableaux}
\begin{ytableau}
	1
\end{ytableau}
\longleftrightarrow
\ytableausetup{smalltableaux}
\begin{ytableau}
	0 \\
	1
\end{ytableau}
$

Now apply rule 2b) to these tableaux: $\alpha = -\frac{1}{\sqrt{2}}$.

\textit{In Gelfand--Tsetlin pattern form:}

The transition is
$
	\bigl(
	\begin{smallmatrix}
		3 & \phantom{=} & 2 \\
		\phantom{=} & 2 & \phantom{=}
	\end{smallmatrix}
	\bigr)
	\longleftrightarrow
	\bigl(
	\begin{smallmatrix}
		3 & \phantom{=} & 3 \\
		\phantom{=} & 3 & \phantom{=}
	\end{smallmatrix}
	\bigr)
$. 

Apply rule 4. 
%The smallest value across both patterns is $2$. 
Subtracting $2$ from every entry in both patterns gives
$
	\bigl(
	\begin{smallmatrix}
		1 & \phantom{=} & 0 \\
		\phantom{=} & 0 & \phantom{=}
	\end{smallmatrix}
	\bigr)
	\longleftrightarrow
	\bigl(
	\begin{smallmatrix}
		1 & \phantom{=} & 1 \\
		\phantom{=} & 1 & \phantom{=}
	\end{smallmatrix}
	\bigr)
$.

Now apply rule 2b) to these patterns: 
%as the bottom number in each pattern is different, we have that 
$\alpha = -\frac{1}{\sqrt{2}}$.

\textbf{Example 3: }
$
\ytableausetup{smalltableaux}
\begin{ytableau}
	0 & 0 & 0 & 0 & 1 & 1 & 1 \\
	1 & 1 & 1 & 1
\end{ytableau}
\longleftrightarrow
\ytableausetup{smalltableaux}
\begin{ytableau}
	0 & 0 & 0 & 0 & 0 & 1 & 1 \\
	1 & 1 & 1 & 1 & 1
\end{ytableau}
$

\textit{In Weyl tableaux form:}

Apply rule 4 to the four shared columns:
$
\ytableausetup{smalltableaux}
\begin{ytableau}
 	1 & 1 & 1
\end{ytableau}
\longleftrightarrow
\ytableausetup{smalltableaux}
\begin{ytableau}
	0 & 1 & 1 \\
	1 
\end{ytableau}
$

Now apply rule 2b) to these tableaux: $\alpha = -\frac{\sqrt{3}}{2}$.

\newpage

\textit{In Gelfand--Tsetlin pattern form:}

The transition is
$
	\bigl(
	\begin{smallmatrix}
		7 & \phantom{=} & 4 \\
		\phantom{=} & 4 & \phantom{=}
	\end{smallmatrix}
	\bigr)
	\longleftrightarrow
	\bigl(
	\begin{smallmatrix}
		7 & \phantom{=} & 5 \\
		\phantom{=} & 5 & \phantom{=}
	\end{smallmatrix}
	\bigr)
$. 

Apply rule 4. 
%The smallest value across both patterns is $4$. 
Subtracting $4$ from every entry in both patterns gives
$
	\bigl(
	\begin{smallmatrix}
		3 & \phantom{=} & 0 \\
		\phantom{=} & 0 & \phantom{=}
	\end{smallmatrix}
	\bigr)
	\longleftrightarrow
	\bigl(
	\begin{smallmatrix}
		3 & \phantom{=} & 1 \\
		\phantom{=} & 1 & \phantom{=}
	\end{smallmatrix}
	\bigr)
$.

Now apply rule 2b) to these patterns: 
%as the bottom number in each pattern is different, we have that 
$\alpha = -\frac{\sqrt{3}}{2}$.

It is easy to verify that we obtain the same transition amplitudes for each example by using Louck's formula, the terms of which are given in (\ref{louckfirst}) and (\ref{loucksecond}).

\section{Examples of the Schur--Weyl Branching Rule} \label{examplesswbranchingrule}

In Sections \ref{mainresults} and \ref{patternrulessection} we have developed all of the theory that we need to be able to give examples of the Schur--Weyl branching rule.

\textbf{Example 1: Left-to-right version}

Consider the following quantum state of $(\mathbb{C}^2)^{\otimes 4}$
\begin{equation} \label{leftrightex1}
\Ket{\lambda = (2,1), 
t =
\ytableausetup{smalltableaux}
\begin{ytableau}
	0 & 1 \\
	1
\end{ytableau}
, 
y = 
\ytableausetup{smalltableaux}
\begin{ytableau}
	1 & 2 \\
	3
\end{ytableau}
}\Ket{0}
\end{equation}

The only possible additions of a box with entry $4$ to the standard Young tableau $y$ such that we form a standard Young tableau $y^{*}$ having at most two rows are given by
$\ytableausetup{smalltableaux}
\begin{ytableau}
	1 & 2 & 4 \\
	3
\end{ytableau}
$
and
$\ytableausetup{smalltableaux}
\begin{ytableau}
	1 & 2 \\
	3 & 4
\end{ytableau}
$.

The only possible 
%additions of $0$ into the 
standard Weyl tableau $t^{*}$ corresponding to these Young tableaux with a box with entry $0$ added to $t$ are given by
$\ytableausetup{smalltableaux}
\begin{ytableau}
	0 & 0 & 1 \\
	1
\end{ytableau}
$
and
$\ytableausetup{smalltableaux}
\begin{ytableau}
	0 & 0 \\
	1 & 1
\end{ytableau}
$
respectively.

%with updated Young tableaux
%$\ytableausetup{smalltableaux}
%\begin{ytableau}
	%1 & 2 & 4 \\
	%3
%\end{ytableau}
%$
%and
%$\ytableausetup{smalltableaux}
%\begin{ytableau}
	%1 & 2 \\
	%3 & 4
%\end{ytableau}
%$
%respectively.

Using the Pattern Rules given in subsection \ref{patternrules}, we see that the transition amplitude for
$
\ytableausetup{smalltableaux}
\begin{ytableau}
	0 & 1 \\
	1
\end{ytableau}
\longleftrightarrow
\ytableausetup{smalltableaux}
\begin{ytableau}
	0 & 0 & 1 \\
	1
\end{ytableau}
$ equals $\frac{1}{\sqrt{2}}$,
and for 
$\ytableausetup{smalltableaux}
\begin{ytableau}
	0 & 1 \\
	1
\end{ytableau}
\longleftrightarrow
\ytableausetup{smalltableaux}
\begin{ytableau}
	0 & 0 \\
	1 & 1
\end{ytableau}
$ it equals $-\frac{1}{\sqrt{2}}$. 
%as calculated from the Pattern Rules given in subsection \ref{patternrules}.

Hence we have that (\ref{leftrightex1}) is equal to
\begin{equation}
\frac{1}{\sqrt{2}}
\Ket{\lambda = (3,1), 
t =
\ytableausetup{smalltableaux}
\begin{ytableau}
	0 & 0 & 1 \\
	1
\end{ytableau}
, 
y = 
\ytableausetup{smalltableaux}
\begin{ytableau}
	1 & 2 & 4 \\
	3
\end{ytableau}
}
-
\frac{1}{\sqrt{2}}
\Ket{\lambda = (2,2), 
t =
\ytableausetup{smalltableaux}
\begin{ytableau}
	0 & 0 \\
	1 & 1
\end{ytableau}
, 
y = 
\ytableausetup{smalltableaux}
\begin{ytableau}
	1 & 2 \\
	3 & 4
\end{ytableau}
}
\end{equation}
expressed in terms of the Schur--Weyl basis of $(\mathbb{C}^2)^{\otimes 4}$.

%\textbf{Right--to--left version}

%Take as input a Schur--Weyl basis state of $(\mathbb{C}^d)^{\otimes n}$. 
%For each $k \in [d]$, form all possible Weyl tableau with a single $k$ removed, of the same shape as the Young tableau with the box labelled by $n$ removed.
%This will give Schur--Weyl basis states of $(\mathbb{C}^d)^{\otimes n-1}$. Let $\ket{k}$ (where $k$ is the entry removed from the input Weyl tableau) be the second register for each such state.
%Form a superposition of these new pairs of quantum registers, with each amplitude given by the transition amplitude between the Weyl tableau given in the first register of the newly formed pair and the input Weyl tableau.
%This is the desired output, as described above.

\textbf{Example 2: Right-to-left version}

Consider the following quantum state of $(\mathbb{C}^2)^{\otimes 3}$
\begin{equation} \label{rightleftex1}
\Ket{\lambda = (2,1), 
t =
\ytableausetup{smalltableaux}
\begin{ytableau}
	0 & 1 \\
	1
\end{ytableau}
, 
y = 
\ytableausetup{smalltableaux}
\begin{ytableau}
	1 & 2 \\
	3
\end{ytableau}
}
\end{equation}
The only possible standard Weyl tableau with a single $0$ removed of the shape
$\ytableausetup{smalltableaux}
\begin{ytableau}
	1 & 2
\end{ytableau}
$ (as the box containing entry $3$ is to be removed from $y$)
is given by
$\ytableausetup{smalltableaux}
\begin{ytableau}
	1 & 1
\end{ytableau}
$.

Similarly, the only possible standard Weyl tableau with a single $1$ removed of the shape
$\ytableausetup{smalltableaux}
\begin{ytableau}
	1 & 2
\end{ytableau}
$
is given by
$\ytableausetup{smalltableaux}
\begin{ytableau}
	0 & 1
\end{ytableau}
$.

As the transition amplitude for 
$\ytableausetup{smalltableaux}
\begin{ytableau}
	1 & 1
\end{ytableau}
\longleftrightarrow
\ytableausetup{smalltableaux}
\begin{ytableau}
	0 & 1 \\
	1
\end{ytableau}
$ is $-\frac{\sqrt{2}}{\sqrt{3}}$
and for 
$\ytableausetup{smalltableaux}
\begin{ytableau}
	0 & 1
\end{ytableau}
\longleftrightarrow
\ytableausetup{smalltableaux}
\begin{ytableau}
	0 & 1 \\
	1
\end{ytableau}
$ is $\frac{1}{\sqrt{3}}$, we have that (\ref{rightleftex1}) is equal to
\begin{equation}
-\frac{\sqrt{2}}{\sqrt{3}}
\Ket{\lambda = (2,0), 
t =
\ytableausetup{smalltableaux}
\begin{ytableau}
	1 & 1
\end{ytableau}
, 
y = 
\ytableausetup{smalltableaux}
\begin{ytableau}
	1 & 2
\end{ytableau}
}\Ket{0}
+
\frac{1}{\sqrt{3}}
\Ket{\lambda = (2,0), 
t =
\ytableausetup{smalltableaux}
\begin{ytableau}
	0 & 1
\end{ytableau}
, 
y = 
\ytableausetup{smalltableaux}
\begin{ytableau}
	1 & 2
\end{ytableau}
}\Ket{1}
\end{equation}
We see that the first register of each state in the superposition is a Schur--Weyl basis state of $(\mathbb{C}^2)^{\otimes 2}$.

\section{The Quantum Schur Transform} \label{schurtransform}

We mentioned at the end of Section \ref{mainresults} that the Schur--Weyl branching rule gives rise to a very simple procedure for performing the Quantum Schur Transform for any $n$ qudits: that is, the unitary change of basis transformation that maps the computational basis of $(\mathbb{C}^d)^{\otimes n}$ to the Schur--Weyl basis of $(\mathbb{C}^d)^{\otimes n}$ (with its adjoint giving the opposite transformation.)

It is now clear to see that we can perform this transformation merely by iteratively applying the left--to--right version of the Schur--Weyl branching rule on the input computational basis state(s), with the adjoint given by iteratively applying the right--to--left version on Schur--Weyl basis state(s).
This procedure is polynomial in $n$ and $d$.

\subsection{Examples of the Quantum Schur Transform with $d=2$}

We give concrete examples of the Quantum Schur Transform for the case $d=2$ in order to display the speed of our method -- in particular, for performing the transformation with $n$ large -- whilst keeping in mind the brevity of our paper.
However, we re-emphasise here for clarity that our method works for any $d$ (and $n$).

\textbf{Example 1}: Express $\ket{0101}$ in the Schur--Weyl basis of $(\mathbb{C}^2)^{\otimes 4}$.

We apply the left-to-right version of the Schur--Weyl branching rule four times, using the Pattern Rules to calculate the transition amplitudes. 
As
\begin{align*}
\Ket{0101} 
	& = 
\Ket{\lambda = (1,0), 
t =
\ytableausetup{smalltableaux}
\begin{ytableau}
	0 
\end{ytableau}
, 
y = 
\ytableausetup{smalltableaux}
\begin{ytableau}
	1 
\end{ytableau}
}\Ket{101} \\
	& =
\frac{1}{\sqrt{2}}
\Ket{\lambda = (2,0), 
t =
\ytableausetup{smalltableaux}
\begin{ytableau}
	0 & 1
\end{ytableau}
, 
y = 
\ytableausetup{smalltableaux}
\begin{ytableau}
	1 & 2
\end{ytableau}
}\Ket{01}
+
\frac{1}{\sqrt{2}}
\Ket{\lambda = (1,1), 
t =
\ytableausetup{smalltableaux}
\begin{ytableau}
	0 \\
	1
\end{ytableau}
, 
y = 
\ytableausetup{smalltableaux}
\begin{ytableau}
	1 \\
	2
\end{ytableau}
}\Ket{01} \\
	& =
\frac{1}{\sqrt{3}}
\Ket{\lambda = (3,0), 
t =
\ytableausetup{smalltableaux}
\begin{ytableau}
	0 & 0 & 1
\end{ytableau}
, 
y = 
\ytableausetup{smalltableaux}
\begin{ytableau}
	1 & 2 & 3
\end{ytableau}
}\Ket{1}
-
\frac{1}{\sqrt{6}}
\Ket{\lambda = (2,1), 
t =
\ytableausetup{smalltableaux}
\begin{ytableau}
	0 & 0 \\
	1
\end{ytableau}
, 
y = 
\ytableausetup{smalltableaux}
\begin{ytableau}
	1 & 2 \\
	3
\end{ytableau}
}\Ket{1} \\
& \phantom{=} +
\frac{1}{\sqrt{2}}
\Ket{\lambda = (2,1), 
t =
\ytableausetup{smalltableaux}
\begin{ytableau}
	0 & 0 \\
	1
\end{ytableau}
, 
y = 
\ytableausetup{smalltableaux}
\begin{ytableau}
	1 & 3 \\
	2
\end{ytableau}
}\Ket{1}
\end{align*}
we have that
\begin{align*}
\Ket{0101}
	& =
\frac{1}{\sqrt{6}}
\Ket{\lambda = (4,0), 
t =
\ytableausetup{smalltableaux}
\begin{ytableau}
	0 & 0 & 1 & 1
\end{ytableau}
, 
y = 
\ytableausetup{smalltableaux}
\begin{ytableau}
	1 & 2 & 3 & 4
\end{ytableau}
}
+ \frac{1}{\sqrt{6}}
\Ket{\lambda = (3,1), 
t =
\ytableausetup{smalltableaux}
\begin{ytableau}
	0 & 0 & 1 \\ 
	1
\end{ytableau}
, 
y = 
\ytableausetup{smalltableaux}
\begin{ytableau}
	1 & 2 & 3 \\
	4
\end{ytableau}
} \\
& \phantom{=} -
\frac{1}{2\sqrt{3}}
\Ket{\lambda = (3,1), 
t =
\ytableausetup{smalltableaux}
\begin{ytableau}
	0 & 0 & 1 \\
	1
\end{ytableau}
, 
y = 
\ytableausetup{smalltableaux}
\begin{ytableau}
	1 & 2 & 4 \\
	3
\end{ytableau}
}
-
\frac{1}{2\sqrt{3}}
\Ket{\lambda = (2,2), 
t =
\ytableausetup{smalltableaux}
\begin{ytableau}
	0 & 0 \\
	1 & 1
\end{ytableau}
, 
y = 
\ytableausetup{smalltableaux}
\begin{ytableau}
	1 & 2 \\
	3 & 4
\end{ytableau}
} \\
& \phantom{=} +
\frac{1}{2}
\Ket{\lambda = (3,1), 
t =
\ytableausetup{smalltableaux}
\begin{ytableau}
	0 & 0 & 1 \\
	1
\end{ytableau}
, 
y = 
\ytableausetup{smalltableaux}
\begin{ytableau}
	1 & 3 & 4 \\
	2
\end{ytableau}
}
+
\frac{1}{2}
\Ket{\lambda = (2,2), 
t =
\ytableausetup{smalltableaux}
\begin{ytableau}
	0 & 0 \\
	1 & 1
\end{ytableau}
, 
y = 
\ytableausetup{smalltableaux}
\begin{ytableau}
	1 & 3 \\
	2 & 4
\end{ytableau}
}
\end{align*}

\textbf{Example 2}: Express
\begin{equation} \label{example2}
\Ket{\lambda = (2,2), 
t =
\ytableausetup{smalltableaux}
\begin{ytableau}
	0 & 0 \\
	1 & 1
\end{ytableau}
, 
y = 
\ytableausetup{smalltableaux}
\begin{ytableau}
	1 & 3 \\
	2 & 4
\end{ytableau}
}
\end{equation}
in the computational basis of $(\mathbb{C}^2)^{\otimes 4}$.

We apply the right-to-left Schur--Weyl branching rule four times, again using the Pattern Rules to calculate the transition amplitudes. 
As equation (\ref{example2}) equals
\begin{align*}
	& \phantom{=}
\frac{1}{\sqrt{2}}
\Ket{\lambda = (2,1), 
t =
\ytableausetup{smalltableaux}
\begin{ytableau}
	0 & 0 \\
	1 
\end{ytableau}
, 
y = 
\ytableausetup{smalltableaux}
\begin{ytableau}
	1 & 3 \\
	2 
\end{ytableau}
}\Ket{1}
-
\frac{1}{\sqrt{2}}
\Ket{\lambda = (2,1), 
t =
\ytableausetup{smalltableaux}
\begin{ytableau}
	0 & 1 \\
	1 
\end{ytableau}
, 
y = 
\ytableausetup{smalltableaux}
\begin{ytableau}
	1 & 3 \\
	2 
\end{ytableau}
}\Ket{0} \\
	& =
\frac{1}{\sqrt{2}}
\Ket{\lambda = (1,1), 
t =
\ytableausetup{smalltableaux}
\begin{ytableau}
	0 \\
	1 
\end{ytableau}
, 
y = 
\ytableausetup{smalltableaux}
\begin{ytableau}
	1 \\
	2 
\end{ytableau}
}[\Ket{01} - \Ket{10}] \\
	& = 
\frac{1}{2}
\Ket{\lambda = (1,0), 
t =
\ytableausetup{smalltableaux}
\begin{ytableau}
	0
\end{ytableau}
, 
y = 
\ytableausetup{smalltableaux}
\begin{ytableau}
	1
\end{ytableau}
}[\Ket{101} - \Ket{110}]
-
\frac{1}{2}
\Ket{\lambda = (1,0), 
t =
\ytableausetup{smalltableaux}
\begin{ytableau}
	1
\end{ytableau}
, 
y = 
\ytableausetup{smalltableaux}
\begin{ytableau}
	1
\end{ytableau}
}[\Ket{001} - \Ket{010}]
\end{align*}
we have that equation (\ref{example2}) equals
\begin{align*}
\frac{1}{2}\Ket{0101} 
- \frac{1}{2}\Ket{0110} 
- \frac{1}{2}\Ket{1001} 
+ \frac{1}{2}\Ket{1010}
\end{align*}

\section{Related Work} \label{relatedwork}

Bacon, Chuang and Harrow were the first to construct the Quantum Schur Transform for the $n$-fold tensor product $(\mathbb{C}^2)^{\otimes n}$ \cite{baconharrowI}, which they extended to $(\mathbb{C}^d)^{\otimes n}$ in \cite{baconharrowII}.
Their approach uses the representation theory of the unitary group, namely by implementing Clebsch-Gordan transforms on a quantum computer and then by iteratively applying these transforms to construct the Quantum Schur Transform.
The quantum algorithm that they present in their extension paper is polynomial in $n, d$ and $\log \frac{1}{\epsilon}$, where $\epsilon$ is the precision.
Whilst our work will result in the same implementation of
%leads to the same procedure for 
the Quantum Schur Transform, it differs significantly in that we use the representation theory of the symmetric group as the theoretical foundation for our approach.
Consequently, we only consider unitary group representations 
%that are appropriate to our problem by applying 
when we apply the Schur--Weyl duality to chains of irreducible representations of the symmetric group that 
%appear naturally in 
form the basis of our approach.

Krovi \cite{Krovi_2019} presented a quantum algorithm for the Quantum Schur Transform that is polynomial in $n, \log d$ and $\log \frac{1}{\epsilon}$.
His approach uses the representation theory of the symmetric group but in a different way to ours, namely by block diagonalising induced representations of the symmetric group known as permutation modules. 

Kaczor and Jakubczyk \cite{kaczor} described a procedure for performing the Quantum Schur Transform that is based on the Schensted insertion and the Robinson--Schensted--Knuth algorithm.
Their approach is based on so-called fundamental shift operators which can be used to find Clebsch--Gordan coefficients for the unitary group. 
This ultimately leads to their being able to calculate the probability amplitudes for the Schur--Weyl basis states appearing in the output of the Quantum Schur Transform.
%We note for the avoidance of any doubt that 
While the method that they use to construct the Schur--Weyl states -- by introducing a directed graph of Gelfand--Tsetlin patterns -- is similar in nature to the left--to--right version of the Schur--Weyl branching rule that we described above,
%develop in the following.
%However, their method 
it differs significantly to ours in that it is derived from the representation theory of the unitary group, and is given without proof, 
whereas our method is derived from the Okounkov--Vershik formulation of the representation theory of the symmetric group, 
which leads not only to a left-to-right version but also to a right-to-left version of the Schur--Weyl branching rule. 
Moreover, we provide a proof of all of our results.

\section{Conclusion} \label{conclusion}

In the preceding sections, we have shown how the Okounkov--Vershik formulation of the representation theory of the symmetric groups $S_n$ can be adapted naturally to the Schur--Weyl duality, resulting in a new way of understanding how the duality can be used to perform the Quantum Schur Transform.
%implementation of the Quantum Schur Transform in $\textbf{WHAT TIME}$ time.

In doing so, we have provided a different theoretical foundation for this transformation from those that have appeared in the literature previously.
%have been shown to exist previously.
In particular, we have constructed a new multigraph called the Schur--Weyl--Young graph, and shown how this corresponds to a branching rule for the Schur--Weyl basis states of the $n$-fold tensor product space $(\mathbb{C}^d)^{\otimes n}$.

We have also found some simple rules for calculating the transition amplitudes between standard Weyl tableaux in adjacent levels of the Schur--Weyl--Young graph for the case $d=2$, which are based entirely on looking at the entries in the Weyl tableaux themselves. We have called these rules the Pattern Rules.
%It would be an interesting challenge to try to generalise the Pattern Rules to any $d$.
It is left as an interesting challenge to the reader to try to generalise the Pattern Rules to any $d$.

Our contributions now make it possible for anyone to change basis between the computational basis of $(\mathbb{C}^d)^{\otimes n}$ and the Schur--Weyl basis $(\mathbb{C}^d)^{\otimes n}$ for any $d, n$ very quickly.
Previously, this had been a slow and painstaking process.
As a result, the attention in quantum computing can now turn to working out which Schur--Weyl states are the most desirable to prepare on a quantum computer, 
%and to determine new quantum algorithms for how they can be used in information processing.
and to construct new quantum algorithms involving their use for the purpose of solving information processing problems.

\section{Acknowledgments}
The author would like to thank his PhD supervisor Professor William J. Knottenbelt 
%for his efforts in securing the funding which has made this research possible
for being generous with his time throughout the author's period of research prior to the publication of this paper.

%The author gratefully acknowledges the Doctoral Scholarship he was awarded under Imperial College London's Department of Computing Applied Research scheme, which has provided the entirety of the funding for this work to be undertaken.
This work was funded by the Doctoral Scholarship for Applied Research which was awarded to the author under Imperial College London's Department of Computing Applied Research scheme.
This work will form part of the author's PhD thesis at Imperial College London.

%ADD REFERENCES
\begin{flushleft}
\nocite{*} %prints all entries of bibliography in References whether cited or not
\bibliographystyle{siam}
\bibliography{references}
\end{flushleft}

%APPENDIX
\newpage
\begin{appendices}

\section{Proof of the Pattern Rules for $d=2$} \label{patternproof}
%\subsection{Amplitude calculations for $d=2$ up to and including $n=6$ using the Pattern Rules}

We claimed in Section \ref{patternrulessection} that the Pattern Rules give the transition amplitude (denoted by $\alpha$ in the following) between standard Weyl tableaux of the following shapes:

	\textbf{Rule 1:} $(n-1, 0) \longleftrightarrow (n,0)$

	\textbf{Rule 2:} $(n-1, 0) \longleftrightarrow (n-1,1)$

	\textbf{Rule 3:} $(n-k-1, k) \longleftrightarrow (n-k,k)$

	\textbf{Rule 4:} $(n-k, k-1) \longleftrightarrow (n-k,k)$

Since these are the only possible transitions between shapes for the case $d=2$ in the Schur--Weyl--Young graph, it is enough to show that our Pattern Rules give the same results as those coming from applying Louck's formula (the terms of which are given in (\ref{louckfirst}) and (\ref{loucksecond})).

We prove this by using the equivalent Gelfand--Tsetlin patterns for each rule.

Given that Louck's formula considers the partial hooks for each entry of a Gelfand--Tsetlin pattern, we define, for convenience, the pattern of partial hooks for the Gelfand--Tsetlin pattern
$$
	\Bigl(
	\begin{smallmatrix}
		m_{1,2}^{n-1} & \phantom{=} & m_{2,2}^{n-1} \\
		\phantom{=} & m_{1,1}^{n-1} & \phantom{=}
	\end{smallmatrix}
	\Bigr)
$$
to be
\begin{equation}
	\Bigl(
	\begin{smallmatrix}
		p_{1,2} & \phantom{=} & p_{2,2} \\
		\phantom{=} & p_{1,1} & \phantom{=}
	\end{smallmatrix}
	\Bigr)
	\coloneqq
	\Bigl(
	\begin{smallmatrix}
		m_{1,2}^{n-1} + 1 & \phantom{=} & m_{2,2}^{n-1} \\
		\phantom{=} & m_{1,1}^{n-1} & \phantom{=}
	\end{smallmatrix}
	\Bigr)
\end{equation}
where $p_{i,j}$ is the partial hook, defined to be equal to $m_{i,j}^{n-1} + j - i$.

We now take each rule in turn.

\textbf{Rule 1:}
We consider the edge between Gelfand--Tsetlin patterns of the form
$$
	\Bigl(
	\begin{smallmatrix}
		n-1 & \phantom{=} & 0 \\
		\phantom{=} & m_{1,1}^{n-1} & \phantom{=}
	\end{smallmatrix}
	\Bigr)
	\longleftrightarrow
	%\text{ --- }
	%\rightarrow
	\Bigl(
	\begin{smallmatrix}
		n & \phantom{=} & 0 \\
		\phantom{=} & m_{1,1}^{n} & \phantom{=}
	\end{smallmatrix}
	\Bigr)
$$

a) If $m_{1,1}^{n-1} = m_{1,1}^{n}$, then, from (\ref{transition}), we have that $n = 2, k = 2$ and $t_2 = 1$, and so,
as
$$
	\Bigl(
	\begin{smallmatrix}
		p_{1,2} & \phantom{=} & p_{2,2} \\
		\phantom{=} & p_{1,1} & \phantom{=}
	\end{smallmatrix}
	\Bigr)
	=
	\Bigl(
	\begin{smallmatrix}
		n & \phantom{=} & 0 \\
		\phantom{=} & m_{1,1}^{n-1} & \phantom{=}
	\end{smallmatrix}
	\Bigr)
$$
we see that Louck's formula gives us that
\begin{equation}
	\alpha 
	= \sqrt{
		\left|
		\frac{
			p_{1,2} - p_{1,1}
		}{p_{1,2} - p_{2,2}}
		\right|
	}
	= \sqrt{
		\frac{
			n - m_{1,1}^{n-1}
		}{n}
	}
	= \sqrt{
		\frac{
			n - m_{1,1}^{n}
		}{n}
	}
\end{equation}
since $m_{1,1}^{n-1} = m_{1,1}^{n}$. This is the same result as (\ref{rule1a}).

b) Else we must have that $m_{1,1}^{n-1} + 1 = m_{1,1}^{n}$, and so, from (\ref{transition}), we have that $n = 2, k = 1, t_1 = 1$ and $t_2 = 1$. 
Hence, as 
$$
	\Bigl(
	\begin{smallmatrix}
		p_{1,2} & \phantom{=} & p_{2,2} \\
		\phantom{=} & p_{1,1} & \phantom{=}
	\end{smallmatrix}
	\Bigr)
	=
	\Bigl(
	\begin{smallmatrix}
		n & \phantom{=} & 0 \\
		\phantom{=} & m_{1,1}^{n-1} & \phantom{=}
	\end{smallmatrix}
	\Bigr)
$$
we see that Louck's formula gives us that
\begin{equation}
	\alpha 
	= \sqrt{
		\left|
		\frac{
			p_{1,1} - p_{2,2} + 1
		}{p_{1,2} - p_{2,2}}
		\right|
	}
	= \sqrt{
		\frac{
			m_{1,1}^{n-1} + 1
		}{n}
	}
	= \sqrt{
		\frac{
			m_{1,1}^{n}
		}{n}
	}
\end{equation}
This is the same result as (\ref{rule1b}).

\textbf{Rule 2:}
We consider the edge between Gelfand--Tsetlin patterns of the form
$$
	\Bigl(
	\begin{smallmatrix}
		n-1 & \phantom{=} & 0 \\
		\phantom{=} & m_{1,1}^{n-1} & \phantom{=}
	\end{smallmatrix}
	\Bigr)
	\longleftrightarrow
	%\text{ --- }
	%\rightarrow
	\Bigl(
	\begin{smallmatrix}
		n-1 & \phantom{=} & 1 \\
		\phantom{=} & m_{1,1}^{n} & \phantom{=}
	\end{smallmatrix}
	\Bigr)
$$

a) If $m_{1,1}^{n-1} = m_{1,1}^{n}$, then, from (\ref{transition}), we have that $n = 2, k = 2$ and $t_2 = 2$, and so,
as
$$
	\Bigl(
	\begin{smallmatrix}
		p_{1,2} & \phantom{=} & p_{2,2} \\
		\phantom{=} & p_{1,1} & \phantom{=}
	\end{smallmatrix}
	\Bigr)
	=
	\Bigl(
	\begin{smallmatrix}
		n & \phantom{=} & 0 \\
		\phantom{=} & m_{1,1}^{n-1} & \phantom{=}
	\end{smallmatrix}
	\Bigr)
$$
we see that Louck's formula gives us that
\begin{equation}
	\alpha 
	= \sqrt{
		\left|
		\frac{
			p_{2,2} - p_{1,1} 
		}{p_{2,2} - p_{1,2}}
		\right|
	}
	= \sqrt{
		\frac{
			m_{1,1}^{n-1} 
		}{n}
	}
	= \sqrt{
		\frac{
			m_{1,1}^{n}
		}{n}
	}
\end{equation}
since $m_{1,1}^{n-1} = m_{1,1}^{n}$. This is the same result as (\ref{rule2a}).

b) Else we must have that $m_{1,1}^{n-1} + 1 = m_{1,1}^{n}$, and so, from (\ref{transition}), we have that $n = 2, k = 1, t_1 = 1$ and $t_2 = 2$. 
Hence, as 
$$
	\Bigl(
	\begin{smallmatrix}
		p_{1,2} & \phantom{=} & p_{2,2} \\
		\phantom{=} & p_{1,1} & \phantom{=}
	\end{smallmatrix}
	\Bigr)
	=
	\Bigl(
	\begin{smallmatrix}
		n & \phantom{=} & 0 \\
		\phantom{=} & m_{1,1}^{n-1} & \phantom{=}
	\end{smallmatrix}
	\Bigr)
$$
we see that Louck's formula gives us that
\begin{equation}
	\alpha 
	= - \sqrt{
		\left|
		\frac{
			p_{1,1} - p_{1,2} + 1
		}{p_{2,2} - p_{1,2}}
		\right|
	}
	= - \sqrt{
		\left|
		\frac{
			m_{1,1}^{n-1} - n + 1
		}{n}
		\right|
	}
	= \sqrt{
		\frac{
			n - m_{1,1}^{n} 
		}{n}
	}
\end{equation}
This is the same result as (\ref{rule2b}).

\textbf{Rule 3:}
We consider the edge between Gelfand--Tsetlin patterns of the form
\begin{equation} \label{gtrule3}
	\Bigl(
	\begin{smallmatrix}
		n-k-1 & \phantom{=} & k \\
		\phantom{=} & m_{1,1}^{n-1} & \phantom{=}
	\end{smallmatrix}
	\Bigr)
	\longleftrightarrow
	%\text{ --- }
	%\rightarrow
	\Bigl(
	\begin{smallmatrix}
		n-k & \phantom{=} & k \\
		\phantom{=} & m_{1,1}^{n} & \phantom{=}
	\end{smallmatrix}
	\Bigr)
\end{equation}

a) If $m_{1,1}^{n-1} = m_{1,1}^{n}$, then, from (\ref{transition}), we have that $n = 2, k = 2$ and $t_2 = 1$, and so,
as
$$
	\Bigl(
	\begin{smallmatrix}
		p_{1,2} & \phantom{=} & p_{2,2} \\
		\phantom{=} & p_{1,1} & \phantom{=}
	\end{smallmatrix}
	\Bigr)
	=
	\Bigl(
	\begin{smallmatrix}
		n-k & \phantom{=} & k \\
		\phantom{=} & m_{1,1}^{n-1} & \phantom{=}
	\end{smallmatrix}
	\Bigr)
$$
we see that Louck's formula gives us that
\begin{equation}
	\alpha 
	= \sqrt{
		\left|
		\frac{
			p_{1,2} - p_{1,1}
		}{p_{1,2} - p_{2,2}}
		\right|
	}
	= \sqrt{
		\left|
		\frac{
			n - k - m_{1,1}^{n-1}
		}{n - k - k}
		\right|
	}
	= \sqrt{
		\left|
		\frac{
			(n - 2k) - (m_{1,1}^{n-1}-k)
		}{(n-2k) - 0}
		\right|
	}
\end{equation}
This is the same result that we would have obtained had our edge been
\begin{equation} \label{rule3ares}
	\Bigl(
	\begin{smallmatrix}
		n-2k-1 & \phantom{=} & 0 \\
		\phantom{=} & m_{1,1}^{n-1} - k & \phantom{=}
	\end{smallmatrix}
	\Bigr)
	\longleftrightarrow
	%\text{ --- }
	%\rightarrow
	\Bigl(
	\begin{smallmatrix}
		n-2k & \phantom{=} & 0 \\
		\phantom{=} & m_{1,1}^{n} - k & \phantom{=}
	\end{smallmatrix}
	\Bigr)
\end{equation}
with $m_{1,1}^{n-1} - k = m_{1,1}^{n} - k$, since, from (\ref{transition}), we have, for (\ref{rule3ares}), that $n = 2, k = 2$ and $t_2 = 1$, and our pattern of partial hooks is
$$
	\Bigl(
	\begin{smallmatrix}
		n-2k & \phantom{=} & 0 \\
		\phantom{=} & m_{1,1}^{n-1} - k & \phantom{=}
	\end{smallmatrix}
	\Bigr)
$$

%which corresponds to the Gelfand--Tsetlin pattern
%$$
%	\Bigl(
%	\begin{smallmatrix}
%		n-2k-1 & \phantom{=} & 0 \\
%		\phantom{=} & m_{1,1}^{n-1} - k & \phantom{=}
%	\end{smallmatrix}
%	\Bigr)
%$$
%and hence to the edge
%$$
%	\Bigl(
%	\begin{smallmatrix}
%		n-2k-1 & \phantom{=} & 0 \\
%		\phantom{=} & m_{1,1}^{n-1} - k & \phantom{=}
%	\end{smallmatrix}
%	\Bigr)
%	\longleftrightarrow
%	%\text{ --- }
%	%\rightarrow
%	\Bigl(
%	\begin{smallmatrix}
%		n-2k & \phantom{=} & 0 \\
%		\phantom{=} & m_{1,1}^{n} - k & \phantom{=}
%	\end{smallmatrix}
%	\Bigr)
%$$
%with $m_{1,1}^{n-1} - k = m_{1,1}^{n} - k$.

Consequently this solution corresponds exactly to subtracting $k$ from every entry of our Gelfand--Tsetlin patterns (\ref{gtrule3}) and applying Rule 1a) to (\ref{rule3ares}).

As we have already verified this rule, it means that Louck's formula and Rule 3a) give the same result.

b) Else we must have that $m_{1,1}^{n-1} + 1 = m_{1,1}^{n}$, and so, from (\ref{transition}), we have that $n = 2, k = 1, t_1 = 1$ and $t_2 = 1$. 
Hence, as 
$$
	\Bigl(
	\begin{smallmatrix}
		p_{1,2} & \phantom{=} & p_{2,2} \\
		\phantom{=} & p_{1,1} & \phantom{=}
	\end{smallmatrix}
	\Bigr)
	=
	\Bigl(
	\begin{smallmatrix}
		n-k & \phantom{=} & k \\
		\phantom{=} & m_{1,1}^{n-1} & \phantom{=}
	\end{smallmatrix}
	\Bigr)
$$
we see that Louck's formula gives us that
\begin{equation}
	\alpha 
	= \sqrt{
		\left|
		\frac{
			p_{1,1} - p_{2,2} + 1
		}{p_{1,2} - p_{2,2}}
		\right|
	}
	= \sqrt{
		\left|
		\frac{
			m_{1,1}^{n-1} - k + 1
		}{n - k - k}
		\right|
	}
	= \sqrt{
		\left|
		\frac{
			(m_{1,1}^{n-1} - k) - 0 + 1
		}{(n-2k) - 0}
		\right|
	}
\end{equation}
This is the same result that we would have obtained had our edge been
\begin{equation} \label{rule3bres}
	\Bigl(
	\begin{smallmatrix}
		n-2k-1 & \phantom{=} & 0 \\
		\phantom{=} & m_{1,1}^{n-1} - k & \phantom{=}
	\end{smallmatrix}
	\Bigr)
	\longleftrightarrow
	%\text{ --- }
	%\rightarrow
	\Bigl(
	\begin{smallmatrix}
		n-2k & \phantom{=} & 0 \\
		\phantom{=} & m_{1,1}^{n} - k & \phantom{=}
	\end{smallmatrix}
	\Bigr)
\end{equation}
with $m_{1,1}^{n-1} - k + 1 = m_{1,1}^{n} - k$, since, from (\ref{transition}), we have, for (\ref{rule3bres}), that $n = 2, k = 1, t_1 = 1$ and $t_2 = 1$, and our pattern of partial hooks is
$$
	\Bigl(
	\begin{smallmatrix}
		n-2k & \phantom{=} & 0 \\
		\phantom{=} & m_{1,1}^{n-1} - k & \phantom{=}
	\end{smallmatrix}
	\Bigr)
$$

%which corresponds to the Gelfand--Tsetlin pattern
%$$
%	\Bigl(
%	\begin{smallmatrix}
%		n-2k-1 & \phantom{=} & 0 \\
%		\phantom{=} & m_{1,1}^{n-1} - k & \phantom{=}
%	\end{smallmatrix}
%	\Bigr)
%$$
%and hence to the edge
%$$
%	\Bigl(
%	\begin{smallmatrix}
%		n-2k-1 & \phantom{=} & 0 \\
%		\phantom{=} & m_{1,1}^{n-1} - k & \phantom{=}
%	\end{smallmatrix}
%	\Bigr)
%	\longleftrightarrow
%	%\text{ --- }
%	%\rightarrow
%	\Bigl(
%	\begin{smallmatrix}
%		n-2k & \phantom{=} & 0 \\
%		\phantom{=} & m_{1,1}^{n} - k & \phantom{=}
%	\end{smallmatrix}
%	\Bigr)
%$$
%with $m_{1,1}^{n-1} - k + \neq m_{1,1}^{n} - k$.

Consequently this solution corresponds exactly to subtracting $k$ from every entry of our Gelfand--Tsetlin patterns (\ref{gtrule3}) and applying Rule 1b) to (\ref{rule3bres}).

As we have already verified this rule, it means that Louck's formula and Rule 3b) give the same result.

\textbf{Rule 4:}
We consider the edge between Gelfand--Tsetlin patterns of the form
\begin{equation} \label{gtrule4}
	\Bigl(
	\begin{smallmatrix}
		n-k & \phantom{=} & k-1 \\
		\phantom{=} & m_{1,1}^{n-1} & \phantom{=}
	\end{smallmatrix}
	\Bigr)
	\longleftrightarrow
	%\text{ --- }
	%\rightarrow
	\Bigl(
	\begin{smallmatrix}
		n-k & \phantom{=} & k \\
		\phantom{=} & m_{1,1}^{n} & \phantom{=}
	\end{smallmatrix}
	\Bigr)
\end{equation}

a) If $m_{1,1}^{n-1} = m_{1,1}^{n}$, then, from (\ref{transition}), we have that $n = 2, k = 2$ and $t_2 = 2$, and so,
as
$$
	\Bigl(
	\begin{smallmatrix}
		p_{1,2} & \phantom{=} & p_{2,2} \\
		\phantom{=} & p_{1,1} & \phantom{=}
	\end{smallmatrix}
	\Bigr)
	=
	\Bigl(
	\begin{smallmatrix}
		n-k+1 & \phantom{=} & k-1 \\
		\phantom{=} & m_{1,1}^{n-1} & \phantom{=}
	\end{smallmatrix}
	\Bigr)
$$
we see that Louck's formula gives us that
\begin{equation}
	\alpha 
	= \sqrt{
		\left|
		\frac{
			p_{2,2} - p_{1,1}
		}{p_{2,2} - p_{1,2}}
		\right|
	}
	= \sqrt{
		\left|
		\frac{
			(k - 1) - m_{1,1}^{n-1}
		}{(k - 1) - (n - k + 1)}
		\right|
	}
	= \sqrt{
		\left|
		\frac{
			0 - (m_{1,1}^{n-1}-k+1)
		}{0 - (n-2k+2)}
		\right|
	}
\end{equation}
This is the same result that we would have obtained had our edge been
\begin{equation} \label{rule4ares}
	\Bigl(
	\begin{smallmatrix}
		n-2k+1 & \phantom{=} & 0 \\
		\phantom{=} & m_{1,1}^{n-1} - k + 1 & \phantom{=}
	\end{smallmatrix}
	\Bigr)
	\longleftrightarrow
	%\text{ --- }
	%\rightarrow
	\Bigl(
	\begin{smallmatrix}
		n-2k+1 & \phantom{=} & 1 \\
		\phantom{=} & m_{1,1}^{n} - k + 1 & \phantom{=}
	\end{smallmatrix}
	\Bigr)
\end{equation}
with $m_{1,1}^{n-1} - k + 1 = m_{1,1}^{n} - k + 1$, since, from (\ref{transition}), we have, for (\ref{rule4ares}), that $n = 2, k = 2$ and $t_2 = 2$, and our pattern of partial hooks is
$$
	\Bigl(
	\begin{smallmatrix}
		n-2k+2 & \phantom{=} & 0 \\
		\phantom{=} & m_{1,1}^{n-1} - k+1 & \phantom{=}
	\end{smallmatrix}
	\Bigr)
$$

Consequently this solution corresponds exactly to subtracting $k-1$ from every entry of our Gelfand--Tsetlin patterns (\ref{gtrule4}) and applying Rule 2a) to (\ref{rule4ares}). 

As we have already verified this rule, it means that Louck's formula and Rule 4a) give the same result.

b) Else we must have that $m_{1,1}^{n-1} + 1 = m_{1,1}^{n}$, and so, from (\ref{transition}), we have that $n = 2, k = 1, t_1 = 1$ and $t_2 = 2$. 
Hence, as 
$$
	\Bigl(
	\begin{smallmatrix}
		p_{1,2} & \phantom{=} & p_{2,2} \\
		\phantom{=} & p_{1,1} & \phantom{=}
	\end{smallmatrix}
	\Bigr)
	=
	\Bigl(
	\begin{smallmatrix}
		n-k+1 & \phantom{=} & k-1 \\
		\phantom{=} & m_{1,1}^{n-1} & \phantom{=}
	\end{smallmatrix}
	\Bigr)
$$
we see that Louck's formula gives us that
\begin{equation}
	\alpha 
	= - \sqrt{
		\left|
		\frac{
			p_{1,1} - p_{1,2} + 1
		}{p_{2,2} - p_{1,2}}
		\right|
	}
	= \sqrt{
		\left|
		\frac{
			m_{1,1}^{n-1} - (n-k+1) + 1
		}{(k-1) - (n-k+1)}
		\right|
	}
	= \sqrt{
		\left|
		\frac{
			(m_{1,1}^{n-1} - k+1) - (n-2k+2) + 1
		}{0 - (n-2k+2)}
		\right|
	}
\end{equation}
This is the same result that we would have obtained had our edge been
\begin{equation} \label{rule4bres}
	\Bigl(
	\begin{smallmatrix}
		n-2k+1 & \phantom{=} & 0 \\
		\phantom{=} & m_{1,1}^{n-1} - k + 1 & \phantom{=}
	\end{smallmatrix}
	\Bigr)
	\longleftrightarrow
	%\text{ --- }
	%\rightarrow
	\Bigl(
	\begin{smallmatrix}
		n-2k+1 & \phantom{=} & 1 \\
		\phantom{=} & m_{1,1}^{n} - k + 1 & \phantom{=}
	\end{smallmatrix}
	\Bigr)
\end{equation}
with $m_{1,1}^{n-1} - k + 2 = m_{1,1}^{n} - k + 1$, since, from (\ref{transition}), we have, for (\ref{rule4bres}), that $n = 2, k = 1, t_1 = 1$ and $t_2 = 2$, and our pattern of partial hooks is
$$
	\Bigl(
	\begin{smallmatrix}
		n-2k+2 & \phantom{=} & 0 \\
		\phantom{=} & m_{1,1}^{n-1} - k + 1 & \phantom{=}
	\end{smallmatrix}
	\Bigr)
$$

Consequently this solution corresponds exactly to subtracting $k-1$ from every entry of our Gelfand--Tsetlin patterns (\ref{gtrule4}) and applying Rule 1b) to (\ref{rule4bres}).

As we have already verified this rule, it means that Louck's formula and Rule 4b) give the same result.

Since we have shown that the Pattern Rules for $d=2$ and Louck's formula give the same transition amplitudes for all standard Weyl tableaux connected by an edge in the Schur--Weyl--Young graph, we have consequently proven our claim.

\end{appendices}

\end{document}